\documentclass[letterpaper, 10 pt, conference]{ieeeconf}  

\IEEEoverridecommandlockouts                              
\usepackage{graphics} 
\usepackage{epsfig} 
\usepackage{amsmath} 
\usepackage{amssymb}  

\usepackage{cite}

\newcommand {\qed}{\hfill\mbox{\rule[0pt]{1.3ex}{1.3ex}}}

\newtheorem{assum}{Assumption}
\newtheorem{lem}{Lemma}
\newtheorem{rem}{Remark}

\title{\LARGE \bf
Faster Response in Bounded-Update-Rate, Discrete-time Networks
using Delayed Self-Reinforcement
}

\author{Santosh Devasia
\thanks{S. Devasia is with Faculty of Mechanical Engineering Department, U. of Washington, Seattle,
WA 08195-2600 USA
        {\tt\small devasia@uw.edu}}%
}

\begin{document}

\maketitle
\thispagestyle{empty}
\pagestyle{empty}

\begin{abstract}

The response speed  of a  network impacts the efficacy of its actions  to external stimuli. 
However, for a given bound on the update rate, the network-response speed is limited  by the need to maintain stability.  
This work  increases the network-response speed without having to increase the update rate by using  delayed self-reinforcement (DSR), 
where each agent uses its already available information from the network to strengthen its  individual update law. 
Example simulation results are presented that  show more than an order of magnitude improvement in the response speed (quantified using the settling time) with the 
proposed DSR approach.

\end{abstract}

\section{Introduction}
\vspace{-0.01in}
Networks are being proposed for control in variety of applications ranging from fleets of self-driving vehicles, 
distributed sensors, swarms of robots and other unmanned aerial and submersible systems. 
Discrete-time models have been studied to capture the response of  such natural  
and engineered networks, e.g.,~\cite{Vicsek_95,Huth_92,Vicsek_95,Jadbabaie_03,Ren_Beard_05,Olfati_Saber_06}
and transportation systems~\cite{Karafyllis_15}.
The performance of these networks is affected by the response speed. 
For example, a fast network can allow rapid propagation of information acquired by one of the agents across the entire network, and impact 
the cohesion of responses  to  external stimuli, e.g., as seen in biological  flocking,~\cite{Attanasi_14}. 
However, the overall   network's response speed (i.e., convergence to  consensus) depends on the update rate at which  each individual agent obtains  information  from other agents connected through the network and changes  its own state. 
The bound on the update-rate can arise due to the time needed for each individual  agent to sense, e.g., the time needed for an ultrasound sensor to ping  and measure distance to a neighbor, 
process information, e.g., time needed for communication and computation, and to react, e.g., due to actuator bandwidth limitations 
(e.g., computational costs as well as time needed for sufficient averaging in noisy environments), and communicate with other agents. 
For a given bound on the update rate (i.e., the smallest time between updates), the  network-response speed is limited  by the need to maintain stability.  
Under such  update-rate-bounded  scenarios, 
the main contributions of this work are (i)~to use delayed self-reinforcement (DSR), 
where each agent uses its current and previously available information to strengthen its  update, and (ii)~to show that such self reinforcement 
can increase the network speed without the need to increase the individual agent's  update  rate.
Similar use of prior update has been used to improve the convergence of gradient-based learning algorithms, and is referred to as the Nesterov's gradient or accelerated gradient method, e.g.,
~\cite{Rumelhart_86,QIAN1999145}.
Moreover, example simulations are used in this article to show that,  without the proposed DSR,  an order-of-magnitude decrease in the settling time requires at-least a corresponding 
order-of-magnitude increase in the update rate. In contrast, the proposed DSR approach achieves more than an order of magnitude improvement in the response speed without 
the need to change the update rate or change the network structure. The impact of this increased response speed, on the ability to maintain relative positions in a formation (without additional control actions) 
 is also illustrated with the simulation results.

The convergence of discrete-time networks has been well studied. Briefly,  it depends on the eigenvalues of the Perron matrix $P$, which maps the current state $I(k)$ to the updated state 
$ I(k+1)   = P I(k)$. 
A variety of methods are available to select the Perron matrix $P$ including the use of neural network methods as in~\cite{Chen_Weisheng_14} and approaches to ensure robust stability, e.g., in~\cite{van_Horssen_16}. 
For example, one choice is to chose the Perron matrix $P$ based on the Laplacian $K$ associated with the underlying graph as
\begin{equation}
I(k+1)   = P I(k) ~ =   
\left[  {\textbf{I}} -\gamma K \right] I(k), 
\label{system_non_source}
\end{equation} 
where  $\gamma$ is the update gain.  
If the underlying graph  is undirected and connected, then convergence to consensus can be achieved provided the update gain $\gamma$ is  
sufficiently small, e.g.,~\cite{Olfati_Murray_07}. 
Robust selection of the distributed controllers for interconnected discrete-time systems has been studied in~\cite{van_Horssen_16}. 
and neural networks have been proposed to select the network~\cite{Chen_Weisheng_14} 
The gain $\gamma$ for a given network Laplacian $K$ can be selected to maximize the convergence rate, i.e.,  the number of updates $N_{T_s}$ needed to achieve convergence to a specified level. 
Note that the network response time $T_s $ is a product of the number of updates $N_{T_s}$ needed and the update time $\delta_t$. 
Alternatively, the scaled Laplacian $\gamma K$ could be optimized for fast convergence, as in~\cite{Carli_08,Boyd_2004}. 
Nevertheless, convergence can  be slow if the number of agent inter-connections is small compared to the number of agents, e.g.,~\cite{Carli_08}. 
Therefore, there is interest in increasing the convergence rate. Previous work has shown that time-varying  connections such as randomized 
interconnections can lead to faster convergence,  e.g.,~\cite{Carli_08}. Another approach is to sequence the update of the agents to improve convergence, e.g.,~\cite{Fanti_15}.
When such time-variations in the graph structure or selection of the graph Laplacian $K$ are not feasible, 
the need to maintain stability 
limits the range of acceptable update gain $\gamma$, and therefore, 
limits the rate of convergence. 
This convergence-rate limitation motivates the proposed effort to develop a new approach to improve the network performance. 

The current investigation on convergence under constraints on the update rate is different from studies that 
seek to ensure convergence under say communication delays~\cite{Olfati_Murray_07,Zhao_Bo_17},  
or communication channel effects~\cite{Xu_Liang_16,Meskin_09}, 
which do not necessarily place bounds on the update rate, but needs 
to be considered when investigating stability. 
Previous works have also investigated the minimal  data rate to ensure stability in terms of the quantization of information transferred during each update~\cite{You_Xie_11}. 
Although, such information communication rates are not considered in this article, such criteria needs to be be met within the minimum time between updates.

The article begins with the problem formulation in Section~\ref{Problem_formulation}, that  briefly reviews the selection of the update gain $\gamma$ to ensure convergence 
and then states the problem of reducing the overall  settling time. The solution approach using DSR is presented in Section~\ref{section_solution}, which also develops stability conditions 
to identify the range of acceptable DSR parameters. Simulations are presented in Section~\ref{Results_and_Discussion} to 
comparatively evaluate the performance improvement with and without DSR. 

and the impact on maintaining a formation. This is followed by the conclusions in Section~V.

\vspace{-0.0in}
\section{Problem formulation}
\label{Problem_formulation}
\vspace{-0.01in}
Agents are assumed to be connected through a network. 
As in current approaches, the state $I_i$  of each agent $i$  is updated using 
information from other agents  connected to agent $i$ through a network, e.g.,~\cite{Olfati_Murray_07}. This section develops properties of this network-based agent update 
to illustrate the limits on the system-response speed (quantified in terms of its settling time $T_s$) when the  update rate cannot be arbitrarily increased. This stability-related limit on the system-response 
speed leads to the   research problem of reducing the settling time $T_s$, stated at the end of this section. 

\subsection{Network definition}
\vspace{-0.01in}
The network is modeled using a graph representation. 
Let the connectivity of the agents be represented by 
a directed graph (digraph) 
${\cal{G}} = \left({\cal{V}}, {\cal{E}}\right)$, e.g., as defined in~\cite{Olfati_Murray_07}, 
with agents represented by nodes $ {\cal{V}}= \left\{ 1, 2, \ldots, {n\!+\!1} \right\}$, $n>1$ 
and edges $ {\cal{E}}   \subseteq {\cal{V}} \times {\cal{V}} $, where each agent $j$ the set of 
neighbors $N_i  \subseteq {\cal{V}} $  of the agent $i$ satisfies  $ j \ne i$ and $(j,i) \in {\cal{E}} $. 
The terms
 $l_{ij}$ of the $(n+1)\times(n+1)$ Laplacian $L$  of the graph ${\cal{G}}$ are real and given by 
\begin{eqnarray}
\label{eq_laplacian_defn}
l_{ij} & =   \left\{ 
\begin{array}{ll}
-w_{ij} < 0, 	& {\mbox{if}} ~ j \in N_i \\
\sum_{m=1}^{n+1} w_{im}, & {\mbox{if}} ~ j = i,  \\
0 & {\mbox{otherwise.}}
\end{array}  
\right.
\end{eqnarray}

Without loss of 
generality,  the source agent is assumed,  
to be the last node, $n+1$. Moreover, all agents in the network should have access to the source agent's state $I_s = I_{n+1}$  through the network, as formalized below.

\vspace{0.01in}
\begin{assum}[Connected to source node]
\label{assum_digraph_properties}
In the following, the digraph ${\cal{G}}$ is assumed  to have a directed path from the source node ${n+1}$ to any  node $ i \in {\cal{V}} \setminus \!{(n+1)}$. 
\end{assum}

Some properties of the Laplacian (L1-L3), used later, are listed below.

{\bf{[L1}}~The $n \times n$ matrix $K$ (the pinned Laplacian), obtained by removing the row and column associated with the source node $n+1$,  
the following partitioning of the Laplacian $L$ is invertible, i.e., 
\begin{eqnarray}
\label{eq_K_eigenvector}
L & = 
\left[
\begin{array}{c|c}
K  &  -B  \\
 \hline
  \star_{1 \times n} & \star_{1 \times 1} 
\end{array}
\right] ~\quad {\mbox{with}} ~~
\det{(K)} \ne 0, 
\end{eqnarray}
and $B$ is an  $n \times 1$ matrix 
\begin{equation}
\label{eq_B_def}
\begin{array}{rl}
B & = [ w_{1,s}, w_{2,s}, \ldots, w_{n,s}]^T
 \\ &  
 ~= [ B_{1}, B_{2}, \ldots, B_{n}]^T , 
\end{array}
\end{equation}
where, under Assumption~\ref{assum_digraph_properties}, the  invertibility of  the pinned Laplacian $K$ expressed in Eq.~\eqref{eq_K_eigenvector}  follows from the Matrix-Tree Theorem in~\cite{tuttle_graph}.

{\bf{[L2]}}~Tthe eigenvalues of $K$ have have strictly-positive, real parts. In particular, from the Gershgorin theorem all the eigenvalues of  the pinned Laplacian $K$ must lie in one of circles centered at $l_{ii} >0$ with radius $ l_{ii}-w_{is}  \in [0, l_{ii}] $  from definition of $l_{ii}$ in Eq.~\eqref{eq_laplacian_defn} and since $w_{is}\ge 0$.  
Moreover, given the invertibility of pinned Laplacian $K$, the   eigenvalues of the pinned Laplacian $K$ cannot be at the origin. Therefore, the eigenvalues must have strictly-positive, real parts (from the Gershgorin theorem of being inside the circles which are on the right hand side of the complex place except for the origin), 
which implies that the eigenvalues $ \lambda_{K,i}$, 
$1 \le i \le n$,  (some of which may be repeated)
\begin{equation}
 \lambda_{K,i}  	~ =  m_{K,i} e^{j \phi_{K,i}}
 \label{eq_lambda_K}
\end{equation}
of the pinned Laplacian $K$ in Eq.~\eqref{eq_K_eigenvector}, 
where $j = \sqrt{-1}$ and 
the magnitudes are ordered as 
\begin{equation}
0 < m_{K,1} ~ \le  m_{K,2}  ~~   \le ~ \ldots ~ \le m_{K,n}, 
\label{eq_positive_mag_ordering_lambda_K}
\end{equation}
have positive real parts, 
i.e., the phase $  \phi_{K,i} $ satisfies   
\begin{equation}
 |  \phi_{K,i}   |   < \frac{\pi}{2}. 
\label{eq_positive_real_parts_lambda_K}
\end{equation}

{\bf{[L3]}}~  
Since each row of the Laplacian $L$ adds to zero, from Eq.~\eqref{eq_laplacian_defn}, 
the $(n+1) \times 1$ vector of ones  ${\textbf{1}}_{n+1}= [1, \ldots, 1]^T$  is a right eigenvector of the Laplacian $L$ with eigenvalue $0$,   
\begin{eqnarray}
\label{eq_first_lambda_eigenvector}
L{\textbf{1}}_{n+1} & = 0  {\textbf{1}}_{n+1} .
\end{eqnarray}
From  Eq.~\eqref{eq_first_lambda_eigenvector}, the partitioning in Eq.~\eqref{eq_K_eigenvector}, and invertibility of $K$, the
product of the inverse of the pinned Laplacian $K$ with $B$ leads to a  $n \times 1$ vector of  ones, i.e., 
\begin{eqnarray}
\label{eq_KinvtimesB}
K ^{-1} B & = {\textbf{1}}_n.
\end{eqnarray}

 \subsection{System description and properties}
\vspace{-0.01in}
Let the network dynamics,   for the non-source agents $I$ represented by the remaining graph ${\cal{G}}\!\setminus\!s$,  be  given by 
\begin{equation}
\begin{array}{rl}
I(k+1) &  = I(k)  -\gamma_t \delta_t K I (k) + {\gamma}_t \delta_t  B  I_s (k)
 \\
&  = I(k)  -{\gamma} K I (k) + {\gamma}   B  I_s (k) \\
& = P I(k) + \gamma B  I_s(k) 
\label{system_non_source}
\end{array}
\end{equation}
where $\delta_t$  is the update rate and  $k$ represents the time instants $t_k = k \delta_t$. 
The  update time $\delta_t$   is considered to be fixed at the fastest possible rate at which each agent can sense, process information and react. 
The  overall update gain 
\begin{equation}
{\gamma} = \gamma_t \delta_t
\label{eq_relations_gain_update_rate}
\end{equation}
needs to be selected to be sufficiently small for stability, i.e., 
all eigenvalues $\lambda_{P,i}$ of the Perron matrix 
$P$ 
\begin{eqnarray}
P &  ~ =  {\textbf{I}}_{n\times n}-\gamma K, 
\label{eq_P_def}
\end{eqnarray}
where $ {\textbf{1}}_{n\times n}$ is the $n\times n$ identity matrix, 
must lie inside the unit circle. 
From the definition of the Perron matrix in Eq.~\eqref{eq_P_def}, 
If $\lambda_{K,i}$ is an eigenvalue of the pinned Laplacian $K$ with 
a corresponding eigenvector $V_{K,i}$, i.e., 
\begin{eqnarray}
K  V_{K,i}   ~ & =    \lambda_{K,i}V_{K,i}, 
\label{eq_stability_condition_lem_Stability_and_Update_gain_pr_1}
\end{eqnarray}
then $ \lambda_{P,i} = 1- \gamma \lambda_{K,i }$ is an eigenvalue of the Perron matrix $P$ for the same eigenvector  $V_{K,i}$, since 
\begin{eqnarray}
P V_{K,i}  ~ =  \left[ {\textbf{I}}_{n\times n}-\gamma K \right]  V_{K,i}
&  
= (1 - \gamma \lambda_{K,i}) V_{K,i}.
\label{eq_stability_condition_lem_Stability_and_Update_gain_pr_2}
\end{eqnarray}
Therefore,  the eigenvalues $\lambda_{P,i}$ of the Perron matrix $P$ in Eq.~\eqref{eq_P_def} are, 
for  $1 \le i \le n$, 
\begin{eqnarray}
\lambda_{P,i}  ~ & =  1 - \gamma_t \delta_t \lambda_{K,i} ~ =  1 - \gamma \lambda_{K,i} . 
\label{Eq_limits_lambdaP_1}
\end{eqnarray}

A sufficiently small selection of the update gain $\gamma$ will stabilize the system in Eq.~\eqref{system_non_source}, e.g., see~\cite{Olfati_Murray_07}.
For a given pinned Laplacian $K$, the range of the acceptable update gain $\gamma$ 
is clarified in the lemma below.

\vspace{0.01in}
\begin{lem}[Perron matrix properties]~ \hfill
\label{lem_Stability_and_Update_gain}
Under Assumption~\ref{assum_digraph_properties}, the network dynamics in Eq.~\eqref{system_non_source}, is stable if and only if 
the update gain $\gamma$ satisfies  
\begin{eqnarray}
0 ~ < \gamma ~ & <  \min_{1\le i \le n}  2   \frac{ \cos{(\phi_{K,i})} }{m_{K,i}}  = \overline\gamma  ~ < ~ \infty .
\label{eq_stability_condition_lem_Stability_and_Update_gain}
\end{eqnarray}
\end{lem}

\vspace{-0.01in}
{\bf{Proof}~}
For stability,  the eigenvalues $\lambda_{P,i}$ of the Perron matrix $P$ need to less than one in magnitude, i.e., for all $1\le i \le n$, 
\begin{eqnarray}
| \lambda_{P,i}|  = |1- \gamma \lambda_{K,i } | ~~ &   < 1.
\label{eq_upper_lower_bound_lambda_P_I}
\end{eqnarray}
Taking squares on both sides results in 
{{
\begin{eqnarray}
\hspace{-0.05in}
|1- \gamma \lambda_{K,i } |^2  & =  \left|1-   \gamma m_{K,i } \cos{(\phi_{K,i})} - j \gamma m_{K,i } \sin{(\phi_{K,i})}  \right|^2  \nonumber \\[0.2em] 
& =  
1-   2 \gamma m_{K,i } \cos{(\phi_{K,i})} + \gamma^2 m_{K,i }^2    < 1  \nonumber 
\end{eqnarray}
}}

\vspace{-0.4in}
\begin{eqnarray}
{\mbox{or}}~~~
-  2 \gamma m_{K,i }  \cos{(\phi_{K,i})}  + \gamma^2 m_{K,i }^2      ~~ & < 0 
\label{condition_proof_lessthanzero}
\end{eqnarray}
leading to the following condition if the update gain satisfies $\gamma > 0 $
\begin{eqnarray}
 0 ~<~  \gamma    ~~ & <  2   \frac{ \cos{(\phi_{K,i})} }{m_{K,i }} . 
 \label{condition_proof_lessthanzero_2}
\end{eqnarray}
Note that the condition in Eq.~\eqref{condition_proof_lessthanzero} is not satisfied if $\gamma=0$. 
Also, if $\gamma < 0 $ then, from Eq.~\eqref{condition_proof_lessthanzero},  $\gamma$ needs to satisfy 
\begin{eqnarray}
 \gamma    ~~ & > 2   \frac{ \cos{(\phi_{K,i})} }{m_{K,i }}   , 
\end{eqnarray}
which is not possible for any negative  $\gamma$ since the right hand side is positive 
for any $i$  from Eqs.~\eqref{eq_positive_mag_ordering_lambda_K} and \eqref{eq_positive_real_parts_lambda_K}.
Therefore, the update gain $\gamma$ needs to satisfy the condition in Eq.~\eqref{eq_stability_condition_lem_Stability_and_Update_gain}. 
Note that the upper bound $\overline\gamma$ in Eq.~\eqref{eq_stability_condition_lem_Stability_and_Update_gain} is finite because 
the magnitude $m_{K,i}$ is nonzero for any $1 \le i \le n$.
\hfill$\qed$
\vspace{0.01in}

\begin{figure}[!ht]
\begin{center}
\includegraphics[width=.75\columnwidth]{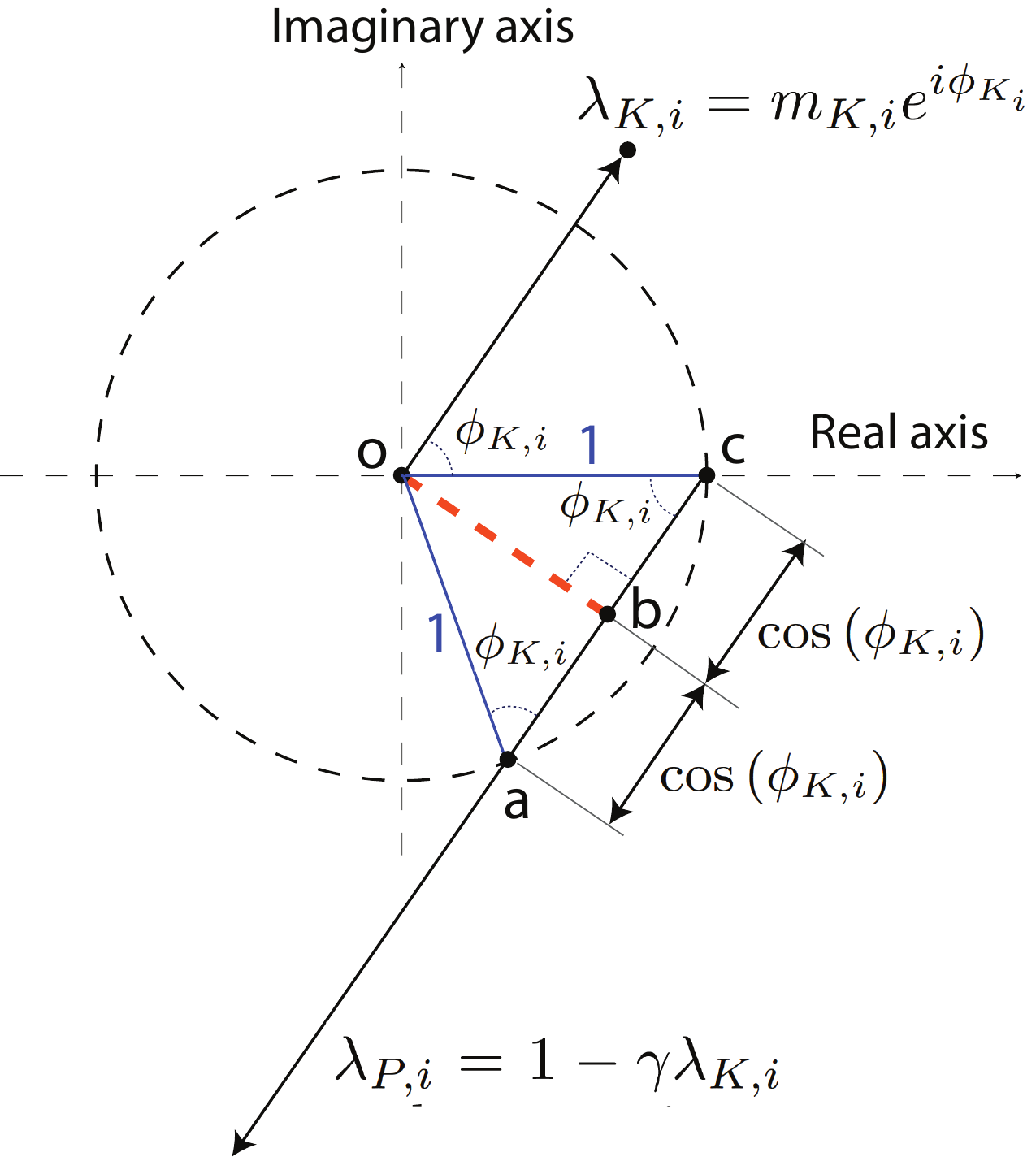}
\vspace{0.01in}
\caption{A geometric interpretation of the result in Lemma~\ref{lem_Stability_and_Update_gain}. 
For stability, the eigenvalue $ \lambda_{P_i} = 1- \gamma  \lambda_{K_i}$   of the Perron matrix P  must be inside the unit circle, i.e.,  the product 
$\gamma \lambda_{K_i}$ must be smaller than the length $d(a,c)= 2 \cos{(\phi_{K,i})} $, as in Eq.~\eqref{condition_proof_lessthanzero_2}. 
}
\label{fig_1_geometry_proof}
\end{center}
\end{figure}

\vspace{0.01in}
\begin{rem}[Geometric interpretation of stability]
A geometric interpretation of the result in Lemma~\ref{lem_Stability_and_Update_gain} is illustrated in Fig.~\ref{fig_1_geometry_proof}. 
The angle $\phi_{K,i}$ is equal to angles $\angle{ocb} $ and $\angle{oab} $. Hence, length $d(b,c) = d(a,b) = \cos{(\lambda_{K,i})} $ where 
$ d(o,c) =1$. 
As a result, for stability, the update gain $\gamma$ should be chosen such the product of the update gain  $\gamma$ and the magnitude 
$ m_{K,i}$ of the eigenvector $\lambda_{K,i}$ should be smaller than length $d(a,c)= 2 \cos{(\phi_{K,i})} $, 
as in Eq.~\eqref{condition_proof_lessthanzero_2}. Moreover, the smallest magnitude of the eigenvalue $\lambda_{P,i}$ over different values of the update gain $\gamma$ is the 
perpendicular distance $d(o,b)$ 
from the origin $o$ to the ray $1-\gamma \lambda_{K_i}$, as illustrated in Fig.~\ref{fig_1_geometry_proof}. This occurs when the product 
$\gamma m_{K,i}$ matches the length $d(b,c)= \cos{(\lambda_{K,i})} $ and is given by $d(o,b)= \sqrt{1 - [\cos{(\phi_{K,i})]^2 } }$.  
\end{rem}

\suppressfloats

\vspace{0.01in}
\begin{lem}[Case with real eigenvalues]~ 
\label{cor_Stability_and_Update_gain_topological_ordering}
Under Assumption~\ref{assum_digraph_properties}, let the  eigenvalues $ \lambda_{K,i}$,
$1 \le i \le n$, 
of the pinned Laplacian $K$ in Eq.~\eqref{eq_K_eigenvector} be real and be ordered as 
\begin{eqnarray}
 \lambda_{K,1} ~ \le  \lambda_{K,2}  ~~   \le ~ \ldots ~ \le \lambda_{K,n} .
\label{eq_positive_ordering_lambda_K}
\end{eqnarray} 
\begin{enumerate}
\item
Then, the eigenvalues of the pinned Laplacian are positive, i.e., 
\begin{eqnarray}
0 <  \lambda_{K,1}, 
\label{eq_positive_ordering_lambda_K_1}
\end{eqnarray}
\item
and the  network dynamics in Eq.~\eqref{system_non_source}, is stable if and only if 
the update gain $\gamma = \gamma_t \delta_t$ satisfies  
\begin{eqnarray}
0 ~ < \gamma ~ & <      \frac{ 2 }{\lambda_{K,n}}  < \infty.
\label{eq_stability_condition_cor_Update_gain}
\end{eqnarray}
\item 
Moreover, if the network gain $\gamma$ satisfies Eq.~\eqref{eq_stability_condition_lem_Stability_and_Update_gain}  then the 
eigenvalues  $\lambda_{P,i}$ of the Perron matrix $P$ in Eq.~\eqref{eq_P_def} satisfy 
\begin{eqnarray}
-1 <  \lambda_{P,n} ~~ \le  \lambda_{P,n-1} ~~\le  \ldots   ~ \le  \lambda_{P,1}  ~~ < 1 .
\label{eq_ordering_P}
\end{eqnarray}
\end{enumerate}
\end{lem}

\vspace{-0.01in}
{\bf{Proof}~}
Statement~1 of the lemma follows 
since 
the phase angle $ \phi_{K,i} =0$ in 
Eq.~\eqref{eq_positive_real_parts_lambda_K} and  $m_{K,1}= \lambda_{K,1}$ in Eq.~\eqref{eq_positive_mag_ordering_lambda_K}. 
The second statement of the lemma follows from Eq.~\eqref{eq_stability_condition_lem_Stability_and_Update_gain} in Lemma~\ref{lem_Stability_and_Update_gain}. 
The last statement of the lemma follows  from Eqs.~\eqref{Eq_limits_lambdaP_1} and ~\eqref{eq_upper_lower_bound_lambda_P_I}, 
 and the ordering in Eq.~\eqref{eq_positive_ordering_lambda_K}. \hfill$\qed$
%
%
 

\vspace{0.01in}
\begin{rem}[Real or acyclic graphs]
\label{Rem_Toplogically_ordered_graphs}
The pinned Laplacian $K$ will have real eigenvalues as in Lemma~\ref{cor_Stability_and_Update_gain_topological_ordering} if it is symmetric or if the associated graph ${\cal{G}}\!\setminus\!s$  is acyclic. 
In an acyclic graph there  is a 
topological ordering of the vertices ${\cal{V}}$ and every graph edge  ${\cal{E}}$ goes from a vertex that is earlier in the ordering to a vertex that is later in the ordering, 
 i.e., all the neighbors $N_i$ of a vertex $i$ are earlier in the ordering. The resulting pinned Laplacian $K$ for an acyclic graph is real-valued and lower triangular, and hence, has real eigenvalues. 
\end{rem}

\vspace{0.01in}
\begin{rem}[Topologically-ordered subgraphs]
\label{Rem_Toplogically_ordered_subgraphs}
A more general case when the pinned Laplacian $K$ has real-valued eigenvalues  is when the 
digraph  ${\cal{G}}\!\setminus\!s$ 
can be partitioned into a 
set of  topologically-ordered subgraphs ${\cal{G}}_i$. Here, the  subgraphs ${\cal{G}}_i$ are distinct (i.e., without shared vertices) where each subgraph is either symmetric or acyclic (topologically ordered)  
with an additional topological ordering of the subgraphs ${\cal{G}}_i$  such that 
all  
graph edges in ${\cal{G}}\!\setminus\!s$   ends in one of the subgraphs, say ${\cal{G}}_i$  and starts: 
(a)~either in the same ending subgraph ${\cal{G}}_i$; or (b) in a subgraph that is earlier than the ending subgraph ${\cal{G}}_i$ in the subgraph ordering. 
In such cases, 
the pinned Laplacian $K$,  associated with the graph  ${\cal{G}}\!\setminus\!s = \bigcup {\cal{G}}_i $   is lower block-triangular, where each block on the diagonal
$K_i$  (associated with $ {\cal{G}}_i $) is real-valued  and either symmetric or lower triangular. Therefore, the eigenvalues of each $K_i$, and therefore,  the eigenvalues of 
$K$  are real. 
\end{rem}

\begin{figure}[!ht]
\begin{center}
\includegraphics[width=.95\columnwidth]{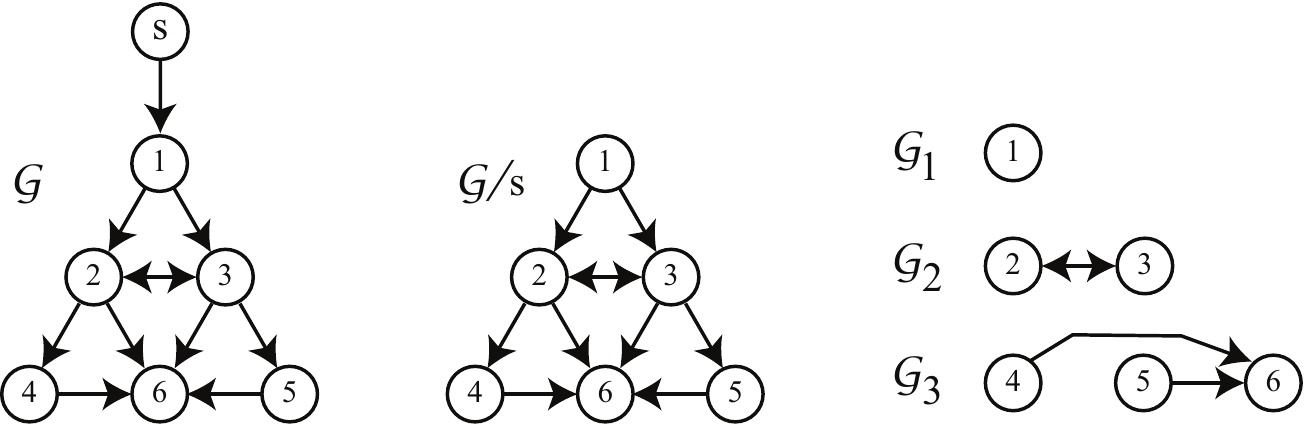}
\vspace{0.01in}
\caption{Example topologically-ordered subgraphs  ${\cal{G}}_1 <  {\cal{G}}_2 <  {\cal{G}}_3$ of graph ${\cal{G}}\!\setminus\!s$ composed of undirected subgraphs ${\cal{G}}_1$, $ {\cal{G}}_2 $ and  ordered acyclic graph $ {\cal{G}}_3$ associated with vertex sets ${\cal{V}}_1 = \left\{ 1 \right\} , {\cal{V}}_2 = \left\{ 2, 3 \right\} $ and  ordered set ${\cal{V}}_3= \left\{ 4 <   5 <  6 \right\} $, 
respectively.  
 All edges in graph ${\cal{G}}\!\setminus\!s$   ends in one of the subgraphs, say ${\cal{G}}_i$  and starts: 
(a)~either in the same ending subgraph ${\cal{G}}_i$; or (b) in a subgraph that is earlier than the ending subgraph ${\cal{G}}_i$ in the subgraph ordering. 
}
\label{fig_2_topological_ordering}
\end{center}
\end{figure}

An example network composed of topologically-ordered subgraphs is provided in Fig.~\ref{fig_2_topological_ordering}. 
The associated  pinned Laplacian $K$,  where the weights $w_{ij}$ in Eq.~\eqref{eq_laplacian_defn} are either $0$ or $1$,  is   
\begin{eqnarray}
K  & = \left[ \begin{array}{cccccc} 
 1     &  0    &  0    &  0     & 0    &  0 \\
  -1    &   2   &  -1   &    0      & 0    &   0 \\
  -1    &  -1    &   2    &   0   &   0    &  0 \\
   0    & -1    &  0     & 1     & 0      &0 \\
   0    &  0   &  -1     & 0  &    1     & 0 \\
   0     &-1   &  -1  &   -1    & -1      & 4
\end{array} \right]
\label{topological_ordering_example_K_2}
\end{eqnarray}
with diagonal blocks  
\begin{eqnarray}
K_1  & = [1],  ~ K_2 =  \left[ \begin{array}{cc}  2 & -1 \\ -1 & 2 \end{array} \right], ~
 K_3 =  \left[ \begin{array}{ccc}  1 & 0 & 0 \\  0 & 1 & 0 \\  -1 & -1 & 4   \end{array} \right] , 
 \label{topological_ordering_example_K_3}
\end{eqnarray} 
associated with the three subgraphs. 
 The eigenvalues of $K$ are then the 
 eigenvalue $1$ of $K_1$, eigenvalues $1, 3$ of $K_2$  and eigenvalues $1, 1, 4$ of $K_3$. 

\vspace{0.01in}
\begin{rem}[Flocking]
The applicability of the results to networks with topologically-ordered subgraphs is important since such ordering can occur 
line-of-sight based networks, e.g., in biological flocking, as well as engineered swarms. 
\end{rem}

 \vspace{0.05in}
An  example where the  graph ${\cal{G}}\!\setminus\!s$  is composed of  an ordered set of subgraphs ${\cal{G}}_1 <  {\cal{G}}_2 <  {\cal{G}}_3$  is shown in Fig.~\ref{fig_2_topological_ordering}, where ${\cal{G}}_1$ and $ {\cal{G}}_2 $ are undirected subgraphs  and  $ {\cal{G}}_3$  is an ordered acyclic graph associated with vertex sets ${\cal{V}}_1 = \left\{ 1 \right\} , {\cal{V}}_2 = \left\{ 2, 3 \right\} $ and  ordered set ${\cal{V}}_3= \left\{ 4 <   5 <  6 \right\} $, 
respectively. 

\begin{eqnarray}
\label{topological_ordering_example_K_2}
K & = 
\left[
\begin{array}{c|c|c} 
\begin{matrix}  1 \end{matrix} & 
\begin{matrix}    0    &  0  \end{matrix} & 
\begin{matrix}   0     & 0    &  0 \end{matrix} \\
\hline 
\begin{matrix}  -1  \\ -1   \end{matrix} & 
\begin{matrix}   2   &  -1  \\  -1    &   2 \end{matrix} & 
\begin{matrix}   0     & 0    &  0  \\  0      & 0    &   0  \end{matrix} \\
\hline
\begin{matrix}  0  \\ 0 \\ 0    \end{matrix} & 
\begin{matrix}  -1    &  0   \\  0   &  -1     \\  -1   &  -1  \end{matrix} & 
\begin{matrix}   1     & 0      &0 \\   0  &    1     & 0 \\   -1    & -1      & 4\end{matrix}
\end{array}
\right] \\[.75em] 
& = 
\left[
\begin{array}{c|c|c} 
K_1  & 
\begin{matrix}    0    &  0  \end{matrix} & 
\begin{matrix}   0     & 0    &  0 \end{matrix} \\
\hline 
\begin{matrix}  -1  \\ -1   \end{matrix} & 
K_2 & 
\begin{matrix}   0     & 0    &  0  \\  0      & 0    &   0  \end{matrix} \\
\hline
\begin{matrix}  0  \\ 0 \\ 0    \end{matrix} & 
\begin{matrix}  -1    &  0   \\  0   &  -1     \\  -1   &  -1  \end{matrix} & 
K_3
\end{array}
\right].
\end{eqnarray}

\suppressfloats

\vspace{-0.01in}
\subsection{Convergence with changes in source}
\vspace{-0.01in}
From stability, the state $I$ of the network (of all non-source agents)  converges to the new source value 
for a step change in the source information $I_s$,  i.e., $I_s(k) = I_d$ for $k > 0$ and zero otherwise. 
Since the eigenvalues of $P$ are inside the unit circle, the solution to Eq.~\eqref{system_non_source} for the step input 
converges 
\begin{eqnarray}
\left[ I(k+1) - I(k)  \right] &  = P^{k}  \left[ I(1) -  I(0)\right]  \rightarrow 0 
\label{Eq_controlled_gen_soln}
\end{eqnarray}
as $ k \rightarrow \infty$.
Therefore, taking the limit as $k \rightarrow \infty$  in Eq.~\eqref{system_non_source}, and from invertibility of the pinned Laplacian $K $ from 
Eq.~\eqref{eq_K_eigenvector}. 
\begin{eqnarray}
I(k)  \rightarrow K^{-1} B  I_d
\label{Eq_controlled_gen_soln_2}
\end{eqnarray}
 as $ k \rightarrow \infty$.  
Then, from Eq.~\eqref{eq_KinvtimesB}, 
the information $I(k)$  at the non-source agents reaches 
the desired information $I_d$ as time step $k$ increases, i.e, 
\begin{eqnarray}
I (k)  &    \rightarrow {\textbf{1}}_n I_d ~~{\mbox{as}}~~  k \rightarrow \infty.
\label{system_non_source_stability}
\end{eqnarray}

\subsection{Limitation on fast settling}
\vspace{-0.01in}
The model in Eq.~\eqref{system_non_source}  can be rewritten as 
\begin{equation}
\begin{array}{rl}
\frac{I(k+1)- I(k)} {\delta_t} &  =   -\gamma_t   K I (k) + {\gamma}_t  B  I_s (k)
\end{array}
\end{equation}
and for a sufficiently-small update time $\delta_t$ it can 
considered as the discrete version of the continuous-time  dynamics 
\begin{equation}
\begin{array}{rl}
\dot{I}(t) &  =   -\gamma_t K I (k) + \gamma_t  B  I_s (k).
\label{system_non_source_contnuous}
\end{array}
\end{equation}
The eigenvalues of $ \gamma_t K$  increase proportionally with $\gamma_t$. Therefore, the settling time $T_s$ of the continuous time system decreases as the gain $\gamma_t$ increases. 
Consequently,  provided the  update time $\delta_t$  is sufficiently small, the discrete-time response of the system in  Eq.~\eqref{system_non_source}  
will be similar to the continuous-time response of the system in Eq.~\eqref{system_non_source_contnuous} and therefore its settling time should 
also decrease as the  gain $\gamma_t$ increases. 
Nevertheless, with a fixed update rate $\delta_t$, the allowable increase in the gain $\gamma_t$ is bounded because  the 
overall update gain $\gamma = \gamma_t \delta_t$ is bounded by the stability condition in Eq.~\eqref{eq_stability_condition_lem_Stability_and_Update_gain} 
of Lemma~\ref{lem_Stability_and_Update_gain}. 
Thus, the smallest possible update time $\delta_t$ limits the fastest possible settling time for a given network.  ~ \hfill \qed

\subsection{The settling-time improvement problem}
\vspace{-0.01in}
The research problem addressed in this article is to   reduce the settling time $T_s$   (from one consensus state to another) 
under step changes in the source value (i.e., improve convergence) where each agent can modify  its update law 
by using already-available information through the network  structure, i.e.,  through the pinned Laplacian $ K$ in the update law in Eq.~\eqref{system_non_source}.

\vspace{0.01in}
\section{Solution using  DSR}
\label{section_solution}
\vspace{-0.01in}
In this section, the proposed delayed self-reinforcement (DSR) approach to address the settling-time-reduction problem is described, 
followed by the development of  stability-based 
bounds on the acceptable range of the DSR parameter.  
The development of such bounds aid in reducing the search space when  optimizing the DSR parameter to minimize the settling time $T_s$. 

\subsection{Proposed approach}
\vspace{-0.01in}
The proposed approach reinforces the update $ I_i(k+1) -I_i(k)$ 
of each non-source agent $i$ with a delayed  and scaled version of the update  $ \beta \left[  I_i(k) -I_i(k-1) \right]$. 
Then, the overall network state $I$ becomes  
\begin{equation}
\begin{array}{rl}
\left[ I(k+1) -I(k) \right] &   = -\gamma  K I(k) + \gamma  B I_s(k)    \\
& ~~~~~ +  \beta \left[ I(k) - I(k-1) \right], 
\label{system_with_source_DSR}
\end{array}
\end{equation}
where $\beta$ is the DSR  gain. 
Therefore,  the  network-update law  in Eq.~\eqref{system_non_source} is modified to   
\begin{equation}
\begin{array}{rl}
 I(k+1) & = I(k)  -\gamma K I(k) + \gamma  B I_s(k)    \\
& ~~~~~ +  \beta \left[ I(k) - I(k-1) \right], 
\label{system_with_source_2}
\end{array}
\end{equation}
which can be  rewritten as 
\begin{equation}
\hat{I}(k+1)    = \hat{P}  \hat{I}(k) + \hat{B} I_s(k)   
\label{system_with_source_2D}
\end{equation}
where 
{\small{
\begin{eqnarray}
\hat{I}(k)  &   = 
\left[ \begin{array}{c}
{I} (k-1)  \\
I(k) 
\end{array}  \right]
, \quad   
%
%
\hat{B}    = \gamma
\left[ \begin{array}{c}
0   \\
B
\end{array}  \right]  \\[0.5em]
\hat{P}  &   = 
\left[ \begin{array}{cc}
0    &   {\textbf{I}}_{n\times n} \\
  - \beta {\textbf{I}}_{n\times n} ~~~  &  \left(\beta {\textbf{I}}_{n\times n} + P \right)  
\end{array}  \right] .
\label{system_with_source_2D_defs}
\end{eqnarray}
}}

\begin{figure}[!ht]
\begin{center}
\includegraphics[width=.65\columnwidth]{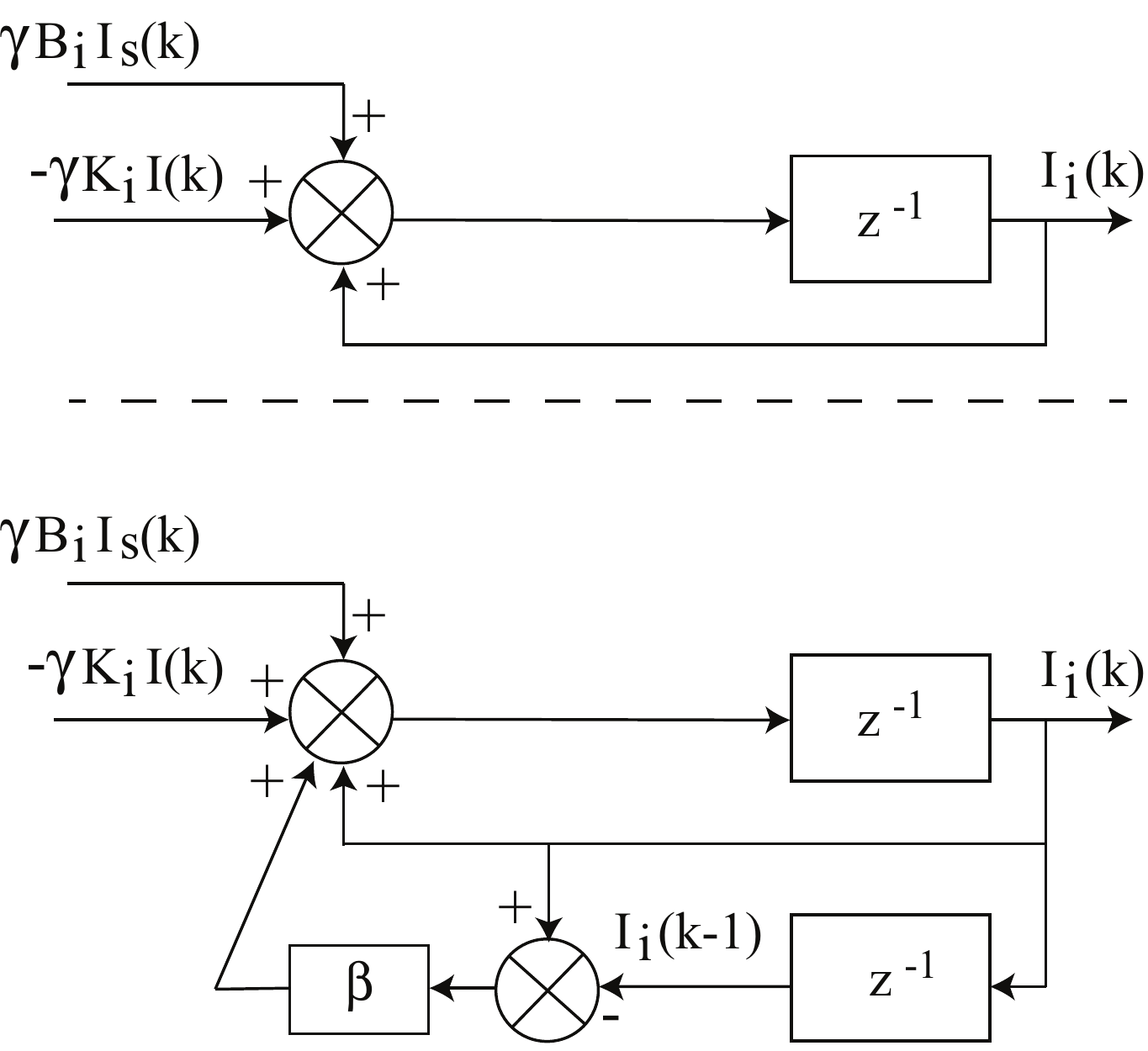}
\vspace{0.01in}
\caption{(Top) Dynamics of agent $i$ for original  networked system without delayed self-reinforcement (DSR). 
(Bottom) Modified dynamics of agent $i$  with delayed self-reinforcement (DSR) 
using the same network information  $K_i Z $ and $B_i z_s $, where the subscript $i$ indicates the $i^{th}$ row of the matrices $K, B$ corresponding to agent $i$ in the 
network. 
}
\label{fig_1_control_implementation}
\end{center}
\end{figure}

\vspace{0.01in}
\begin{rem}[Network information used by DSR]
\label{Rem_Toplogically_ordered_subgraphs}
The  additional information needed to implement the DSR   $ \beta [ I(k)-I(k-1)]$ is available already 
to each agent $i$, i.e.,  the state values  $I_i(k) $and $I_i(k-1)$ at time instant $k$, where  the subscript $i$ indicates the $i^{th}$ agent. 
DSR, however, requires each agent to store a delayed version $I_i(k-1)$ of its current  state $I_i(k)$ and to have knowledge of the DSR gain $\beta$, 
as illustrated in Fig.~\ref{fig_1_control_implementation}.  ~ \hfill \qed
\end{rem}

\suppressfloats

The use of DSR can only improve the performance. 
If the origin $I=0$ of the dynamics without the DSR in Eq.~\eqref{system_non_source} 
is stable, then there is some open interval $(\underline{\beta}, \overline{\beta})$ that contains zero  where the origin  $I=0$ is stable  
with the proposed DSR approach   because 
the eigenvalues of the modified Perron matrix $\hat{P}$ vary continuously with the DSR gain $\beta$. 
Hence the optimal  DSR gain could decrease, but never increase, 
the settling time $T_s$ when compared to the case without  DSR.

\vspace{0.01in}
\subsection{Stability conditions on the DSR gain $\beta$}
\vspace{-0.01in}
The DSR gain  $\beta$ should be selected to reduce the settling time of the system. 
Since the DSR gain $\beta$ is a scalar parameter, numerical search methods could be used to optimally select it. 
Nevertheless, it is helpful to have a bounded search space. 
One approach is to restrict the search space to avoid regions where the 
modified network dynamics in Eq.~\eqref{system_with_source_2D} is expected to be unstable. 
Therefore, the conditions on the DSR gain $\beta$ for network stability  are investigated below.

\vspace{0.01in}
\begin{lem}[Stability with DSR] 
\label{Lemma_DSR_stability}
Let the origin of the system~\eqref{system_non_source} without DSR, i.e.,   $\beta=0$ be stable.
\begin{enumerate}
\item 
Then, the origin of the network dynamics 
with DSR in Eq.~\eqref{system_with_source_2D} is unstable if 
the DSR gain $\beta$ satisfies 
\begin{eqnarray}
| \beta |  ~~ > ~~1.
\label{Eq_bound_beta_stability_DSR}
\end{eqnarray}
\item 
Moreover, if the eigenvalues of the pinned Laplacian are real and ordered as in 
Eq.~\eqref{eq_positive_ordering_lambda_K}, then the origin of the 
network dynamics 
with DSR in Eq.~\eqref{system_with_source_2D} is stable if and only if 
the DSR gain $\beta$ satisfies 
\begin{eqnarray}
- \left[ 1 -\frac{1}{2} \gamma\lambda_{K,n} \right]   ~~  < ~~\beta   ~~ < ~~1.
\label{Eq_bound_beta_stability_DSR_real_eig}
\end{eqnarray}
\end{enumerate}
\end{lem}

\vspace{-0.01in}
{\bf{Proof}~}
The eigenvalues $\lambda_{\hat{P}}$ and the associated eigenvectors  $V_{\hat{P}}$   of the modified Perron matrix $\hat{P}$  satisfy 
\begin{eqnarray}
\lambda_{\hat{P}} V_{\hat{P}}~ = 
\lambda_{\hat{P}} 
\left[ \begin{array}{c}
V_{\hat{P},t}  \\
V_{\hat{P},b}
\end{array}  \right] 
 &   = 
 \hat{P} 
 \left[ \begin{array}{c}
V_{\hat{P},t}  \\
V_{\hat{P},b}
\end{array}  \right] 
\end{eqnarray}
or 
\begin{equation}
\begin{array}{rl}
\lambda_{\hat{P}} 
V_{\hat{P},t} 
& = V_{\hat{P},b} \\
\lambda_{\hat{P}} 
V_{\hat{P},b} 
 &   = 
 -\beta V_{\hat{P},t}  + \left(\beta I + P \right)  V_{\hat{P},b} 
\end{array}
\end{equation}
that can be combined as 
\begin{eqnarray}
\lambda_{\hat{P}} ^2
V_{\hat{P},t} 
 &   = 
 -\beta V_{\hat{P},t}  + \lambda_{\hat{P}}   \left(\beta I + P \right)  V_{\hat{P},t} 
\end{eqnarray}
leading to 
\begin{equation}
\begin{array}{rl}
\left[  \lambda_{\hat{P}} ^2   + \beta  - \beta \lambda_{\hat{P}}    \right]
V_{\hat{P},t} 
 &   = 
 \lambda_{\hat{P}} \left[  P  \right]   V_{\hat{P},t} \\
 & = 
\lambda_{\hat{P}} \left[  {\textbf{I}}_{n\times n}-\gamma K \right]   V_{\hat{P},t} .
\end{array}
\label{eq_eig_value_diagonal_pf}
\end{equation}
Let the  pinned Laplacian $K$, be similar to the matrix $K_J$ in the real-valued Jordan form, where  
\begin{equation}
K_J = T_K^{-1} K T_K
\label{Jordan_K_eq}
\end{equation}
where 
$T_K $ is invertible, e.g.,~\cite{Ortega}.
Then, Eq.~\eqref{eq_eig_value_diagonal_pf} can be rewritten as 
\begin{eqnarray}
\left[  \lambda_{\hat{P}} ^2   + \beta  - \beta \lambda_{\hat{P}}    \right]
V_{J,\hat{P},t} 
 & = 
\lambda_{\hat{P}}   \left[  {\textbf{I}}_{n\times n} -\gamma   K_J    \right]   V_{J,\hat{P},t}  
\label{eq_eig_value_diagonal_pf_3}
\end{eqnarray}
where 
$V_{J,\hat{P},t}    =   T_K^{-1} V_{\hat{P},t} $.
Then, due to the block-triangular form of the matrix $K_{J}$,  
the eigenvalues $\lambda_{\hat{P},i}  $ of the 
modified Perron matrix $\hat{P}$ 
are related to the eigenvalues 
$\lambda_{K,i} = m_{K,i} e^{j \phi_{K,i}} $  in Eq.~\eqref{eq_lambda_K} of the pinned Laplacian $K$ as 
\begin{eqnarray}
\hspace{-0.2in}
\det \left\{  \left[ 
\lambda_{\hat{P},i} ^2   + \beta  - \beta \lambda_{\hat{P},i}   -\lambda_{\hat{P},i} 
\right] {\textbf{I}}_{n\times n}    + \lambda_{\hat{P},i} \gamma K_{J,i} 
\right\}= 0 
\label{eq_det_eig_expression}
\end{eqnarray}
where the diagonal  block $K_{J,i}$  of the real-valued matrix $K_J$  is given by 
{{
\begin{eqnarray*}
K_{J,i}  & =   \left[  \begin{array}{cc}   m_{K,i} \cos(\phi_{K,i})   & m_{K,i} \sin(\phi_{K,i})  \\    - m_{K,i} \sin(\phi_{K,i}) &   m_{K,i} \cos(\phi_{K,i})    \end{array} \right] 
 \\ & = 
\left[  \begin{array}{cc}   a_i  &  b_i  \\    -b_i  &   a_i \end{array} \right] .
\end{eqnarray*}
}}
The eigenvalue Eq.~\eqref{eq_det_eig_expression} can be rewritten as 
\begin{eqnarray}
\hspace{-0.2in} \left[\lambda_{\hat{P},i} ^2   + \beta  - \beta \lambda_{\hat{P},i}   -\lambda_{\hat{P},i}  +a_i \lambda_{\hat{P},i} \gamma
\right]^2  + \lambda_{\hat{P},i}^2 \gamma^2 b_i^2   = 0 
\label{eq_det_eig_expression_2}
\end{eqnarray}
or
\begin{eqnarray}
\left[ \sum_{k=0}^{4} a_{k,i} \lambda_{\hat{P},i} ^k  \right]
&  = 0 
\label{eq_det_eig_expression_3}
\end{eqnarray}
where $a_{4,i}=1$ and 
$a_{4,0}= \beta^2$, which leads to the necessary condition $ a_{4,i} > |a_{4,0}| $ for the eigenvalue to satisfy $|\lambda_{\hat{P},i}| < 1 $  from the Jury test. 
This results in the first statement of the lemma. 

If the eigenvalues of the pinned Laplacian are real, then $ a_i  = \lambda_{K,i}$ and $ b_i =0$ in Eq.~\eqref{eq_det_eig_expression_2}, and the 
eigenvalues $\lambda_{\hat{P}}$  of the modified Perron matrix $\hat{P}$ are given by 
\begin{eqnarray}
\hspace{-0.2in} 
{\cal{P}} \left(  \lambda_{\hat{P},i} \right) ~ = \lambda_{\hat{P},i} ^2  ~+    \left(\gamma \lambda_{K,i} - \beta    -1 \right) \lambda_{\hat{P},i}    ~+ \beta .
  ~~= 0 
\label{eq_det_eig_expression_4}
\end{eqnarray}
Then, from the Jury test, for the eigenvalue to satisfy $|\lambda_{\hat{P},i}| < 1 $, 
\begin{equation}
\begin{array}{rl}
|{\cal{P}}(0)|  < 1, \quad 
{\cal{P}}(1)   > 0 , \quad 
{\cal{P}}(-1)   > 0 , 
\label{eq_det_eig_expression_5}
\end{array}
\end{equation}
which 
results in 
\begin{equation}
\begin{array}{rl}
| \beta |   < 1 , \quad 
\gamma \lambda_{K,i}   > 0 , \quad 
 \beta     > - \left[ 1   -\frac{1}{2} \gamma\lambda_{K,i} \right] .
\label{eq_det_eig_expression_5_22}
\end{array}
\end{equation}
From Eq.~\eqref{eq_stability_condition_cor_Update_gain}, $\gamma \lambda_{K,i}   > 0$ and 
\begin{equation}
\begin{array}{rl}
-1 < - \left[ 1   -\frac{1}{2} \gamma\lambda_{K,i} \right]  & < 0, 
\label{eq_det_eig_expression_5_23} 
\end{array}
\end{equation}
which along with the ordering in Eq.~\eqref{eq_positive_ordering_lambda_K} leads   to the second statement of the lemma 
from Eq.~\eqref{eq_det_eig_expression_5_22}.  \hfill ~\qed

\subsection{Selection of DSR gain}
\vspace{-0.01in}
The optimal value $\beta^*$ for the smallest settling time can be found through a numerical search over the range where the network with DSR is stable as identified in Lemma~\ref{Lemma_DSR_stability}.
An analytical approximation-based approach, based on the continuous-time approximation of the discrete dynamics, is described below 
to select the DSR gain for the case when the eigenvalues of the pinned Laplacian $K$ are real. Moreover, this approximation aids in understanding the potential settling-time improvements 
that can be anticipated with the proposed DSR method.

In this subsection, the eigenvalues of the pinned Laplacian are assumed to be real valued. 
With DSR, the discrete-time system~\eqref{system_with_source_2} can be 
rewritten as 
\begin{equation}
\begin{array}{rl}
\beta \left\{ \left[ I (k+1) - I(k)  \right] - \left[ I (k) - I(k-1)  \right]   \right\} & \\
+ 
(1-\beta) \left[ I(k+1) - I(k)  \right]  & \\
= -\gamma_t \delta_t K I(k) + \gamma_t \delta_t B I_s(k) , 
\end{array}
\label{Eq_DSR_Rearrangement}
\end{equation}
which can be  approximated, 
when the update interval $\delta_t$ is small compared to the dominant network response, as
\begin{equation}
\beta \delta_t^2 \ddot{I}(t)   +(1-\beta) \delta_t \dot{I} (t) = -\gamma_t \delta_t  K I(t) + \gamma_t  \delta_t  B  I_s(t)   , 
\label{Eq_approximation_DSR_continuous}
\end{equation}
which matches the continuous-time system in Eq.~\eqref{system_non_source_contnuous} if 
the DSR gain $\beta$ is zero.

\begin{rem}[Momentum term]
The use of a non-zero DSR gain $\beta$ results in a mass-like term in the approximation in Eq.~\eqref{Eq_approximation_DSR_continuous}. 
Therefore the delayed reinforcement term  $  \left [I_i (k)  - I_i (k-1) \right]$  in Eq.~\eqref{system_with_source_2} is  referred to as the momentum-term in 
gradient-based learning algorithms, 
e.g.,~\cite{QIAN1999145}. ~ \hfill \qed
\end{rem}

With DSR, i.e., a nonzero DSR gain $\beta$, the above equation can be rewritten as 
\begin{equation}
 \ddot{I}(t)   +\frac{(1-\beta)}{\beta \delta_t} \dot{I} (t) = -\frac{\gamma_t}{\beta \delta_t} K I(t) +\frac{\gamma_t}{\beta \delta_t} B I_s(t). 
\label{Eq_approximation_DSR_continuous}
\end{equation}

To consider the impact of the different modes, let the pinned Laplacian $K$ be similar to the matrix $K_J$ in the Jordan form, 
as in Eq.~\eqref{Jordan_K_eq}, 
where  
the diagonal  terms of the real-valued matrix $K_J$ 
are the eigenvalues $\left\{ \lambda_{K,i} \right\}_{i=1}^{n}$   of matrix $K$~\cite{Ortega}. 
Note that the  multiplicity of each eigenvalue $ \lambda_{K,i} $ can be more than one. 
With the transformation into modal coordinates, 
\begin{eqnarray}
I(k) &  =  T_K I_J (k) 
\label{system_non_source_transformation_22}
\end{eqnarray}
the
network dynamics in Eq.~\eqref{Eq_approximation_DSR_continuous} 
can be rewritten as
\begin{equation}
 \ddot{I}_J(t)   +\frac{(1-\beta)}{\beta \delta_t} \dot{I}_J (t) = -\frac{\gamma_t}{\beta \delta_t} K_J I_J(t) +\frac{\gamma_t}{\beta \delta_t} B_J I_s(t).
\label{Eq_approximation_DSR_continuous_modal}
\end{equation}
where $B_J = T_K^{-1} B$. Therefore, for each  pole $s_i$ of the approximate continuous-time dynamics in Eq.~\eqref{system_non_source_contnuous} without DSR, i.e., 
\begin{equation}
\begin{array}{rl}
s_i + \gamma_t \lambda_{K,i} & = 0
\label{pole_system_non_source_continuous}
\end{array}
\end{equation}
the corresponding poles of the approximate continuous-time dynamics in Eq.~\eqref{Eq_approximation_DSR_continuous_modal} with DSR are given by the roots  of 
\begin{equation}
\begin{array}{rl}
s^2 + \frac{(1-\beta)}{\beta \delta_t} s  +\frac{\gamma_t  \lambda_{K,i}}{\beta \delta_t}    & = 0 
\label{poles_approximation_DSR_continuous_modal}
\end{array}
\end{equation}
or
\begin{equation}
\begin{array}{rl}
s^2 + 2 \zeta_i \omega_i s  + \omega_i^2  & = 0 
\label{poles_approximation_DSR_continuous_modal_2}
\end{array}
\end{equation}
where 
\begin{equation}
\begin{array}{rl}
 \omega_i  ~~ = \sqrt{ \frac{\gamma_t \lambda_{K,i} }{\beta \delta_t}    }, & ~~\quad 
 2 \zeta_i \omega_i ~~ = \frac{(1-\beta)}{\beta \delta_t}.
\label{poles_approximation_DSR_continuous_modal_3}
\end{array}
\end{equation}

Without DSR, let the settling time $T_{s,i}$ associated with eigenvalue $ \lambda_{K,i} $  be estimated from Eq.~\eqref{pole_system_non_source_continuous} 
as  a multiple of the time constant, 
the inverse of the distance of the eigenvalue from the imaginary axis in the complex plane, 
(and potential effects of eigenvalue multiplicity are neglected), 
i.e., 
\begin{eqnarray}
T_{s,i}    & \approx  \frac{4}{ | s_i | }  =    \frac{4}{ \gamma_t \lambda_{K,i} }.
\label{TSi_no_DSR}
\end{eqnarray}
Critical  damping for the corresponding eigenvalue with DSR (corresponding to eigenvalue $ \lambda_{K,i} $) occurs when 
damping $\zeta_i=1$, 
i.e., 
\begin{eqnarray}
\zeta_i = \frac{(1-\beta)}{2 \beta \delta_t}\sqrt{  \frac{\beta \delta_t} {\gamma_t  \lambda_{K,i}}   }  = 1 
\label{TSi_with_DSR_2}
\end{eqnarray}
with solution $\beta^* < 1$ given by 
\begin{eqnarray}
 \beta^* & = (1+2    \gamma_t   \delta_t  \lambda_{K,i})  -\sqrt{(1+2   \gamma_t   \delta_t   \lambda_{K,i})^2    -1}  \\
 & = \left(1+ \frac{8\delta_t}{T_{s,i} } \right)  -\sqrt{ \left(1+\frac{8 \delta_t}{T_{s,i}}  \right)^2    -1}.
\label{TSi_with_DSR_2_6}
\end{eqnarray}
In this critically-damped case, the  corresponding settling time $\hat{T}_{s,i}$ can be approximated, 
from Eqs.~\eqref{poles_approximation_DSR_continuous_modal_3} and \eqref{TSi_no_DSR}, as 
(again, potential effects of eigenvalue multiplicity are neglected) 
\begin{eqnarray}
\hat{T}_{s,i}    & \approx  \frac{6}{\zeta_i \omega_i} ~ =   \frac{6}{\omega_i}    ~= 6 \sqrt{ \frac{\beta^* \delta_t} {\gamma_t \lambda_{K,i} }   }  
~ =  
3 \sqrt{ {\beta^*  \delta_t   T_{s,i}}   } .
\label{TSi_with_DSR_2_7}
\end{eqnarray}

\begin{rem}[Reduction in settling time with DSR]
Let the pole $s_i$ of the approximate continuous-time dynamics in Eq.~\eqref{system_non_source_contnuous} be the dominant dynamics (e.g., closest to the imaginary axis of the complex plane) without DSR. 
Then, the settling time $\hat{T}_{s,i}$ with DSR can be substantially smaller than the settling time $T_{s,i}$ without DSR, if 
the update time $\delta_t$ is small and if the settling time without DSR is large, i.e,  $T_{s,i} \gg 1$, 
as seen from Eq.~\eqref{TSi_with_DSR_2_7}. ~ \hfill \qed
\end{rem}

\section{Results and discussion}
\label{Results_and_Discussion}
\vspace{-0.01in}
The step response of an example system, with and without DSR, are comparatively evaluated. Moreover, the impact of using DSR  on the response of the networked system is illustrated 
when the networks state is the turn angle during turn maneuvers of a formation of agents.

\subsection{Example networked system for simulation}
\vspace{-0.01in}
The networked system used in simulation consists of $n=31$ non-source agents connected in a ring, with the leader (agent connected to the source) selected as $i^*=16$, which is 
connected to the source agent  $i=n+1$. The  neighbors of each non-source agent $i$ are given by 
\begin{equation}
\begin{array}{rll}
N_i  = & \{i-1, i+1\}  & \forall   ~ 2 \le i \le n-1  ~ \& ~ i \ne i^* \\
= & \{n, 2\}  & {\mbox{if}}~~ i =1 \\ 
= & \{n-1, 1\}  & {\mbox{if}}~~ i =n \\
= & \{ i-1, i+1, n+1  \}  & {\mbox{if}}~~ i =i^* 
\end{array}
\end{equation}
and weights $w_{ij}=1$ if agent $j$ is a neighbor of agent $i$, i.e., $ j \in N_i $ and zero otherwise. 
The eigenvalues $ \lambda_{K,i}$ 
of the pinned Laplacian $K$ in Eq.~\eqref{eq_K_eigenvector} are real, positive and can be ordered as in Eq.~\eqref{eq_positive_ordering_lambda_K}
with the smallest one $  \lambda_{K,1} =  0.0081$ and the largest one $\lambda_{K,n} =  4.2361$.

\subsection{Selection of update gain}
\begin{figure}[!h]
\begin{center}
    \begin{tabular}{@{}c@{}}
    \includegraphics[width=0.9\columnwidth]{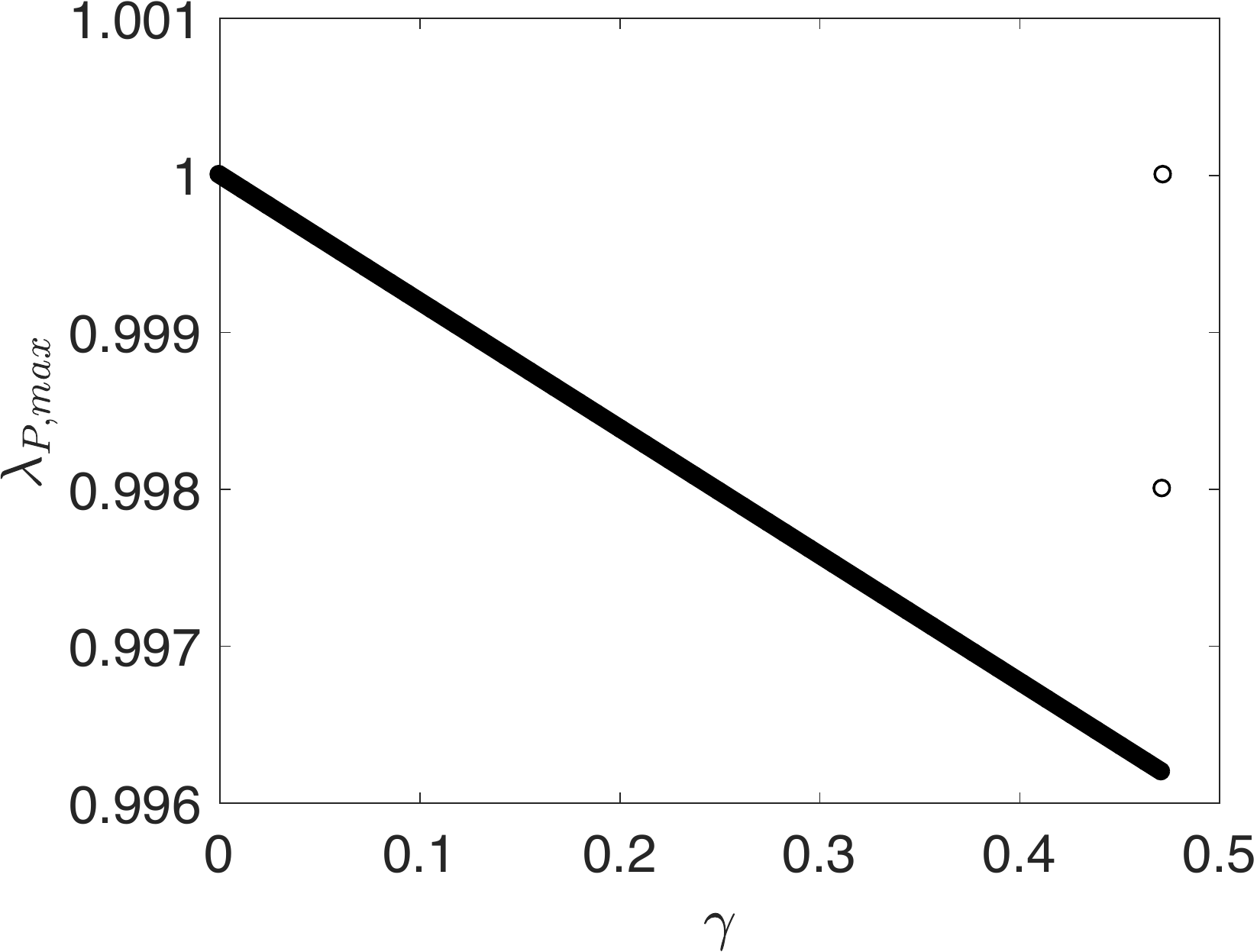}  \\ 
        \includegraphics[width=0.9\columnwidth]{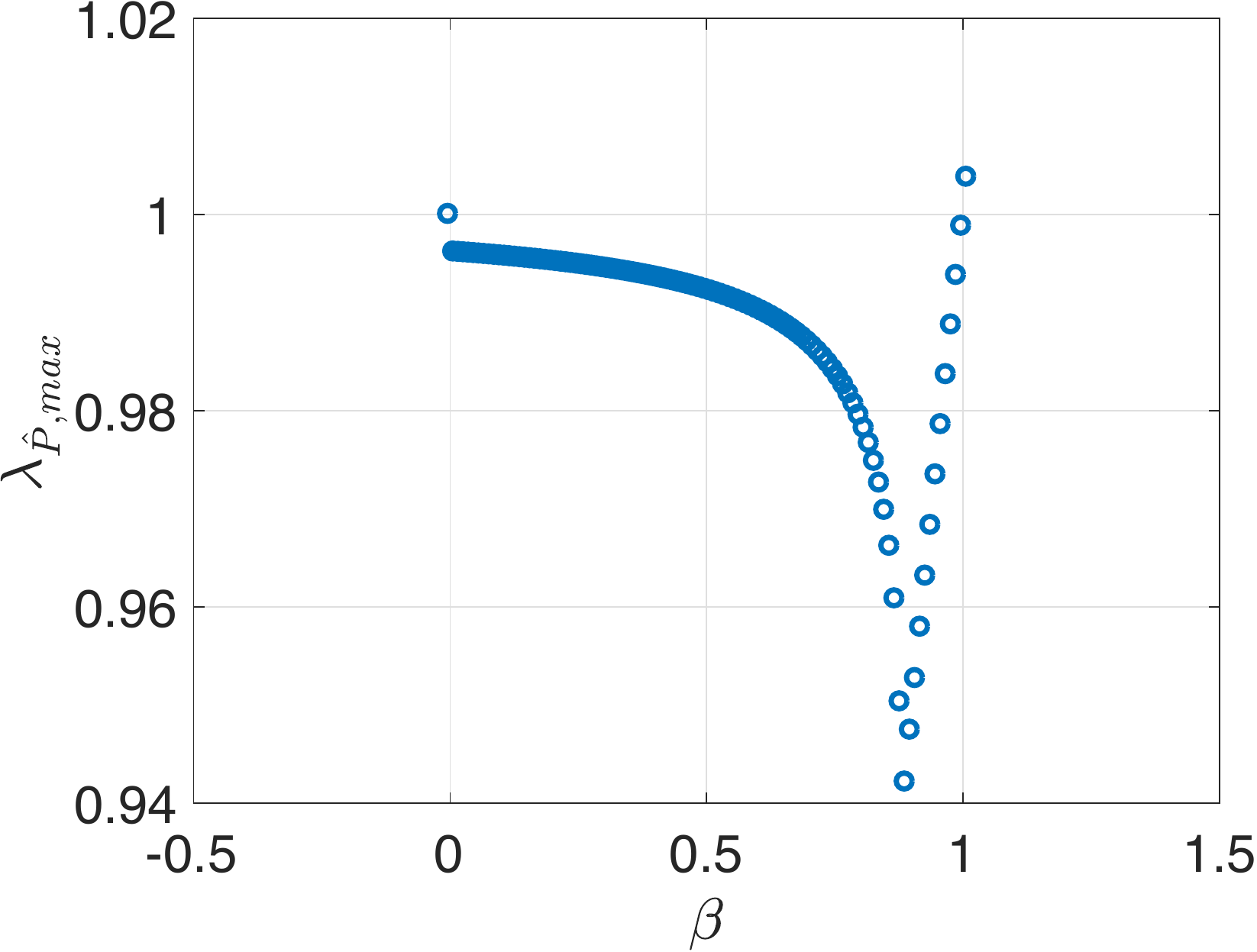} 
  \end{tabular}
   \end{center}
\caption{
(Top) Effect of overall update gain $\gamma $ on the maximum magnitude eigenvalue $\lambda_{P,max} = \max_i |\lambda_{P,i}|$
of the Perron matrix $P$, which is minimized at  $\gamma  = 0.47119$. (Bottom) 
Effect of DSR  gain $\beta $ on the maximum magnitude eigenvalue $\lambda_{\hat{P},max} $ of the modified Perron matrix $\hat{P}$.
}
\label{fig_3_gamma_effect}
\end{figure}

\vspace{-0.01in}
From Lemma~\ref{cor_Stability_and_Update_gain_topological_ordering}, the maximum value of the overall update gain $\gamma $ is 
given from Eq.~\eqref{eq_stability_condition_cor_Update_gain} as 
\begin{eqnarray}
0 ~ < \gamma ~ & <      \frac{ 2 }{\lambda_{K,n}}   =  \overline{\gamma}  ~\approx  0.47214.
\label{exp_stability_condition_cor_Update_gain}
\end{eqnarray}
This is a tight bound on stability, as seen in Fig.~\ref{fig_3_gamma_effect}, the system becomes unstable at $\overline{\gamma}$ since the 
maximum magnitude of the  eigenvalues  
 \begin{eqnarray}
 \lambda_{P,max} ({\gamma})= \max_i |\lambda_{P,i} ({\gamma}) | 
 \end{eqnarray}
 becomes one with $\gamma = \overline{\gamma} $, i.e., 
 \begin{eqnarray} 
 \lambda_{P,max} (\overline{\gamma}) = |1 -\overline{\gamma} \lambda_{K,n} |  = 1.
 \end{eqnarray}
 In the following, the overall update gain $\gamma$ is chosen 
to be $0.471$, which is close to the value where the maximum magnitude eigenvalue $\lambda_{P,max}(\gamma)$ is the 
smallest for fast settling as seen in Fig.~\ref{fig_3_gamma_effect}.

\subsection{Step response without DSR}
\begin{figure}[!h]
\begin{center}
    \begin{tabular}{@{}cc@{}}
    \includegraphics[width=0.9\columnwidth]{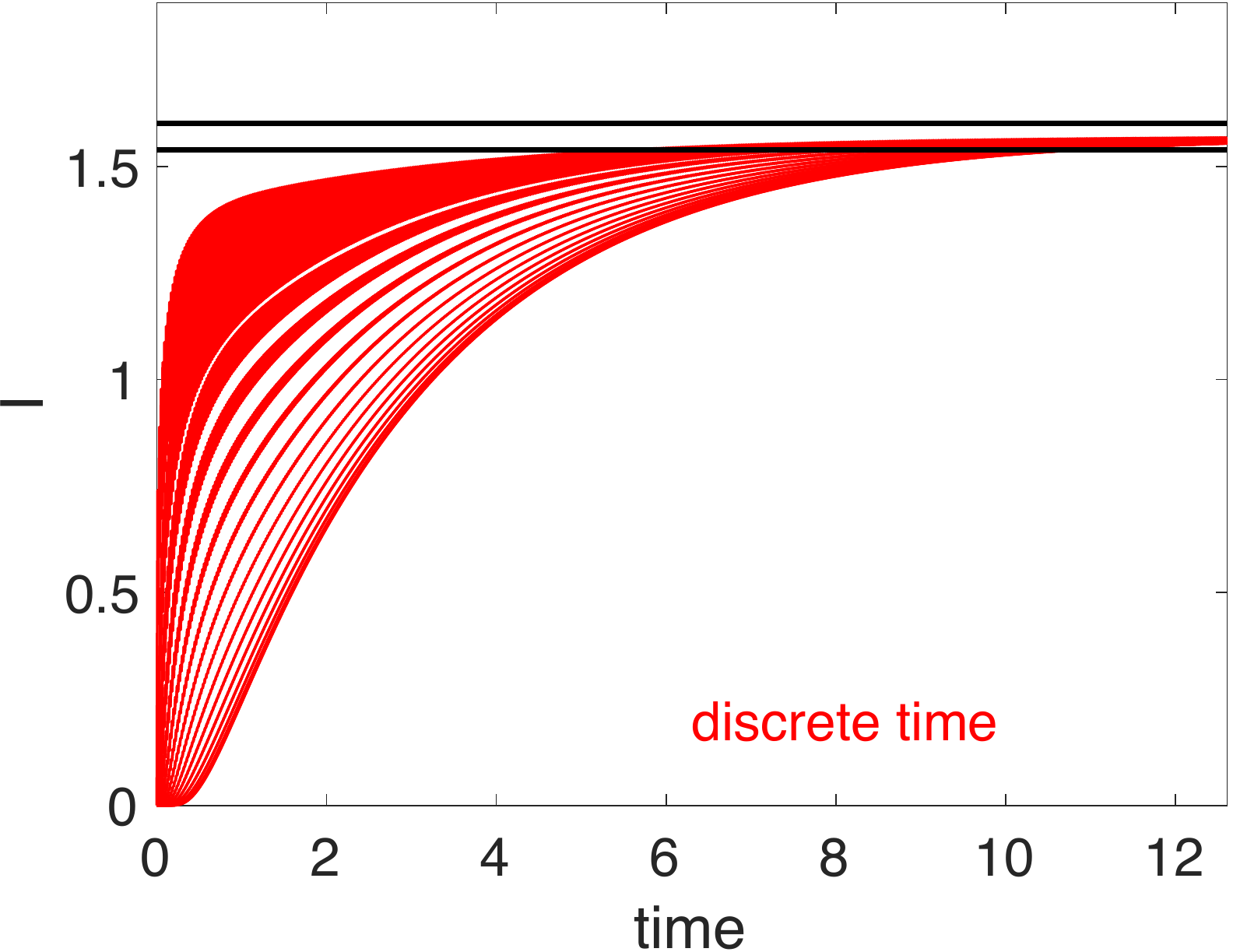}  \\
     \includegraphics[width=0.9\columnwidth]{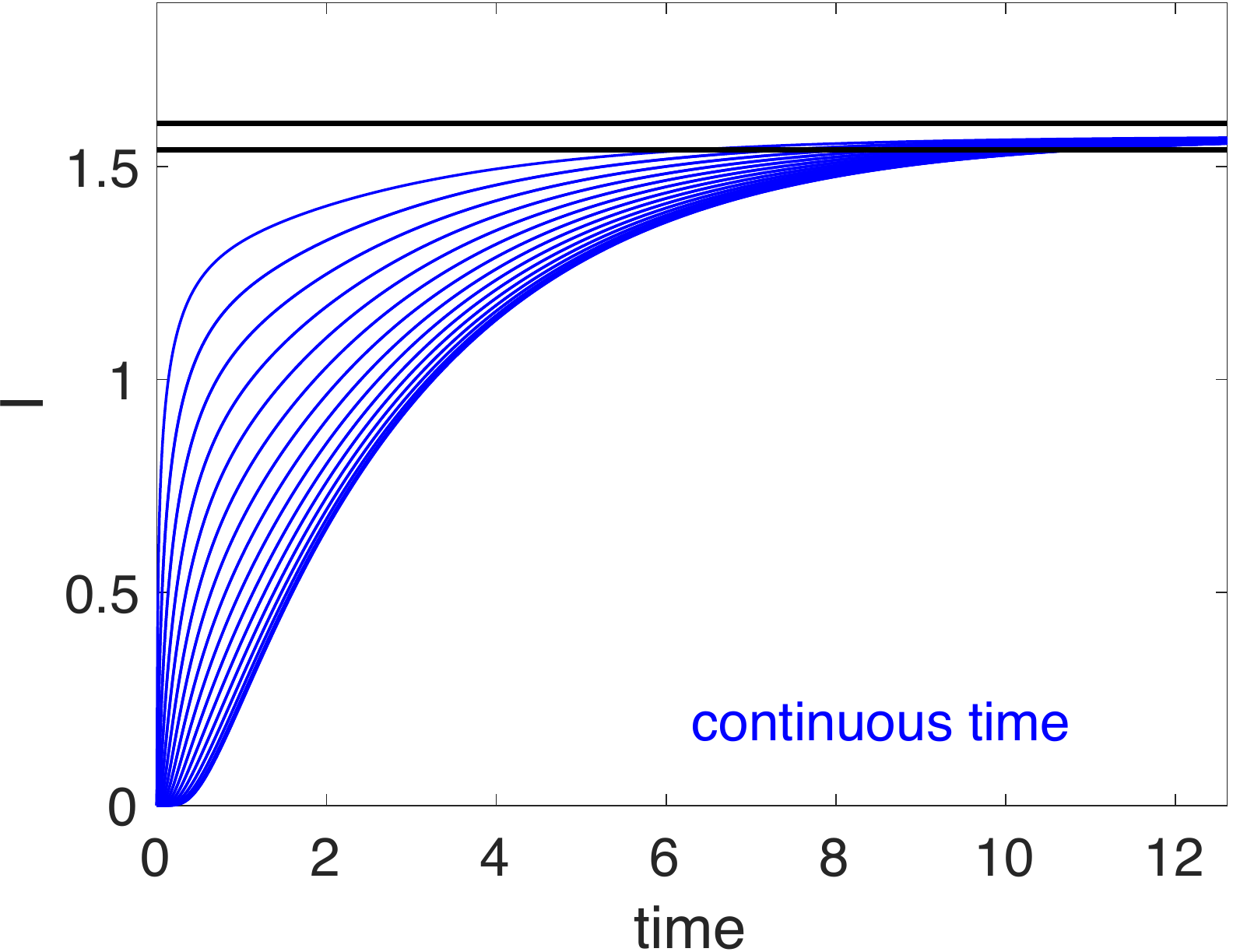}  
  \end{tabular}
   \end{center}
\caption{(Top)~Response of discrete-time network with optimal update gain $\gamma$ to a step input $I_s$ of magnitude $\pi/2$ and update time of $\delta_t=0.01$ s.  
(Bottom)~Similar step response of the approximate continuous-time network in Eq.~\eqref{system_non_source_contnuous} with optimal update gain $\gamma$. The horizontal black lines 
represent $\pm2\%$ deviation from the final value of $\pi/2$.
}
\label{fig_4_step_response_con_dis}
\end{figure}

\vspace{-0.01in}
With the  update gain $\gamma = 0.471$, the  response of the system with a step input of magnitude 
$\pi/2$ and update time $\delta_t=0.01$~s is shown in Fig.~\ref{fig_4_step_response_con_dis}. 
The settling time $T_s$, defined as the time needed for the response to reach and stay within $2\%$ of the final value of $\pi/2$, is $T_s=12.04$~s, which matches the settling time $T_{s,c}=12.07$~s from the 
continuous-time network  approximation in Eq.~\eqref{system_non_source_contnuous} with the same optimal update gain $\gamma$.
Note that these settling times are close to the predicted settling time of $T_{s,1}  = 10.5$~s from Eq.~\eqref{TSi_no_DSR} of the approximated continuous-time dynamics Eq.~\eqref{system_non_source_contnuous} using  the smallest eigenvalue 
$  \lambda_{K,1} =  0.0081$ and $\gamma = 0.471$. The deviation in the predicted settling time with the continuous-time approximation is to be expected since several other eigenvalues are close to $  \lambda_{K,1}$.
Without DSR, the above settling time of $T_s=12.04$~s is the fastest expected settling time for the linear discrete-time network with the fixed update time of $0.01$~s.

\subsection{Step response with DSR}
\vspace{-0.01in}
The settling time  can be faster with the use of DSR. The 
network with DSR is expected to remain stable for DSR gain $\beta$ between 
$-0.0024$ and $1$ from  Eq.~\eqref{Eq_bound_beta_stability_DSR_real_eig} in Lemma~\ref{Lemma_DSR_stability}. 
The variation of the maximum magnitude eigenvalue $\lambda_{\hat{P},max} =  \max_i\{\lambda_{\hat{P},i} \} $ of the 
modified Perron matrix $\hat{P}$ with different DSR gain $\beta$ (at $0.01$ increments) is shown in Fig.~\ref{fig_3_gamma_effect}. Note that the 
maximum magnitude eigenvalue $\lambda_{\hat{P},max} $ 
becomes one at the these boundary values $-0.0024,1 $ of the DSR gain $\beta$ indicating the tightness of 
the range estimate for stability in Eq.~\eqref{Eq_bound_beta_stability_DSR_real_eig}.
The value of the DSR gain $\beta$ was selected to minimize the maximum magnitude of eigenvalues $\lambda_{\hat{P},max} $ of the modified Perron matrix $\hat{P}$ 
(over the computed $\beta$ values), since a larger maximum magnitude eigenvalue $\lambda_{\hat{P},max} $ tends to result in slower settling. 
The  maximum magnitude of eigenvalues $\lambda_{\hat{P},max} $ was minimized at DSR gain  $\beta  = 0.8876$.  This optimal DSR gain $\beta$ 
is close to the approximation-based prediction 
for critical damping of the eigenvalue associated with the smallest eigenvalue $\lambda_{K,1}$ of the pinned Laplacian $K$ from Eq.~\eqref{TSi_with_DSR_2_6} 
\begin{eqnarray}
\hspace{-0.1in}  \beta^* & = (1+2    \gamma \lambda_{K,1})  -\sqrt{(1+2   \gamma  \lambda_{K,1})^2    -1}   = 0.8840, 
 \label{TSi_with_DSR_example}
\end{eqnarray}
that results in a predicted settling time, from Eq.~\eqref{TSi_with_DSR_2_7}, of 
\begin{eqnarray}
\hat{T}_{s,i}    & \approx   6 \sqrt{ \frac{\beta^* \delta_t} {\gamma_t \lambda_{K,1} }   }   = 0.9149 s.
\label{TSi_with_DSR_2_7_example}
\end{eqnarray}

\begin{figure}[!h]
\begin{center}
    \begin{tabular}{@{}c@{}}
    \includegraphics[width=0.99\columnwidth]{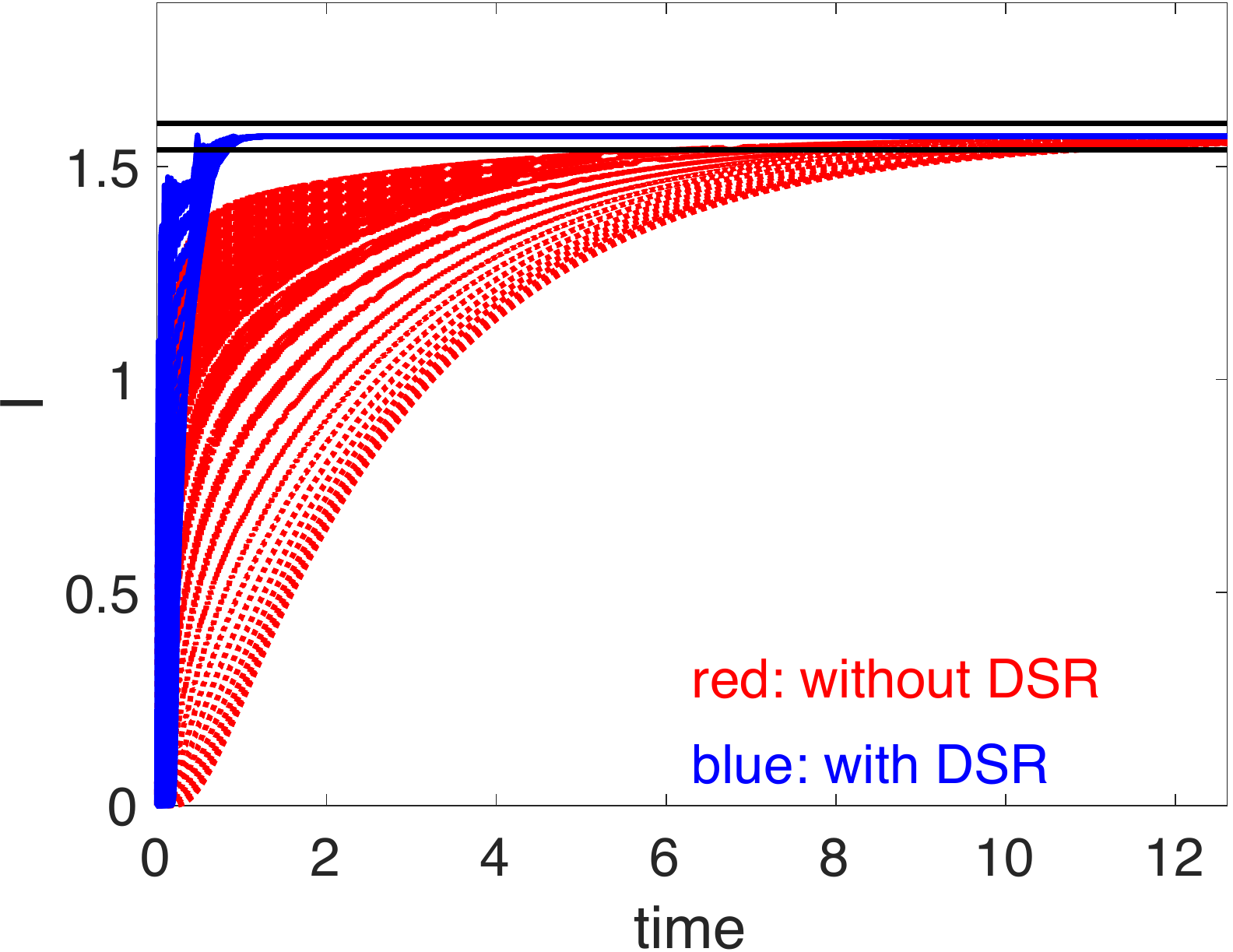} 
  \end{tabular}
   \end{center}
\caption{
Comparison of step response with and without DSR. The DSR gain  was $\beta  = 0.8876$.
}
\label{fig_6_step_responses}
\end{figure}

The step response of the discrete-time system with and without DSR is compared in Fig.~\ref{fig_6_step_responses}. 
The settling time with DSR is $0.90$~s, which is significantly small compared to the settling time of $12.04$~s without DSR. 
Moreover, the achieved settling time of  $0.90$~s is close to that predicted with the approximation-based analysis of $0.9149$~s in Eq.~\eqref{TSi_with_DSR_2_7_example}.
Thus, the proposed DSR approach enables faster settling time  (more than an order-of-magnitude improvement) without the need to change the update rate ($\delta_t =0.01$~s) 
of the system. 

\vspace{0.01in}
\begin{rem}[Smaller settling time without  DSR]
The system without the DSR also would have a faster settling time of $T_s=0.90$~s  if the update gain $\gamma_t$ was selected to be larger by 
the desired reduction in the settling time, i.e., $13.38 = 12.04/0.90$ and if the update time $\delta_t$ was smaller 
by the same factor $13.38$. This results in the same overall update gain $\gamma = \gamma_t \delta_t$, and requires the same number of discrete-time update 
steps (as in Fig.~\ref{fig_4_step_response_con_dis}) 
for settling to within 2\% of the final value, 
but the time interval between the updates is smaller by $13.38$.  ~ \hfill \qed
\end{rem}

\subsection{Comparison with second-order discrete-event dynamics}
\vspace{-0.01in}
Based on the approximate continuous-time 
analysis of the discrete-time system in Eq.~\eqref{Eq_approximation_DSR_continuous}, a faster response can be anticipated if 
each agent's response is second-order. However, this can lead to instability when the update time $\delta_t$ is not changed because the 
effective stiffness of the  system  in Eq.~\eqref{Eq_approximation_DSR_continuous} is higher by $1/{\beta\delta_t}$  when compared to the original 
system approximation in Eq.~\eqref{system_non_source_contnuous}. 

To illustrate, based on Eq.~\eqref{Eq_approximation_DSR_continuous}, let the second-order discrete-time dynamics, 
with an update time of $\tilde{\delta}_t$, be given by
\begin{equation}
\begin{array}{rl}
\tilde{I}(k+1) &   =  \tilde{I}(k)  + \tilde{\delta}_t \tilde{K}  \tilde{I}(k) + \tilde{\delta}_t  \tilde{B} I_s(k)    \\  
& =\tilde{P}  \tilde{I}(k) + \tilde{\delta}_t  \tilde{B} I_s(k)
\end{array}
\label{system_with_source_2D_tilde}
\end{equation}
where 
{{
\begin{eqnarray}
\tilde{I}(k)  &   = 
\left[ \begin{array}{c}
I(k) \\
\dot{I} (k)  
\end{array}  \right]
, \quad   
\tilde{B}    =  \frac{\gamma_t}{\beta \delta_t}
\left[ \begin{array}{c}
0   \\
B
\end{array}  \right]  \\[0.5em]
\tilde{K}  &   = 
\left[ \begin{array}{cc}
0    &   {\textbf{I}}_{n\times n}  \\
   -\frac{\gamma_t}{\beta \delta_t} K  ~~~  &  -\frac{(1-\beta)}{\beta \delta_t} {\textbf{I}}_{n\times n} 
\end{array}  \right]  
\\[0.5em]
\tilde{P}  &   = 
\left[ \begin{array}{cc}
{\textbf{I}}_{n\times n}   &  \tilde{\delta}_t {\textbf{I}}_{n\times n}  \\
   -\frac{\gamma_t\tilde{\delta}_t}{\beta \delta_t} K  ~~~  &  \left(1-\frac{(1-\beta)\tilde{\delta}_t}{\beta \delta_t}\right) {\textbf{I}}_{n\times n}
\end{array}  \right]  .
\label{system_with_source_2D_defs_tilde}
\end{eqnarray}
}}

\begin{figure}[!h]
\begin{center}
    \begin{tabular}{@{}c@{}}
    \includegraphics[width=0.9\columnwidth]{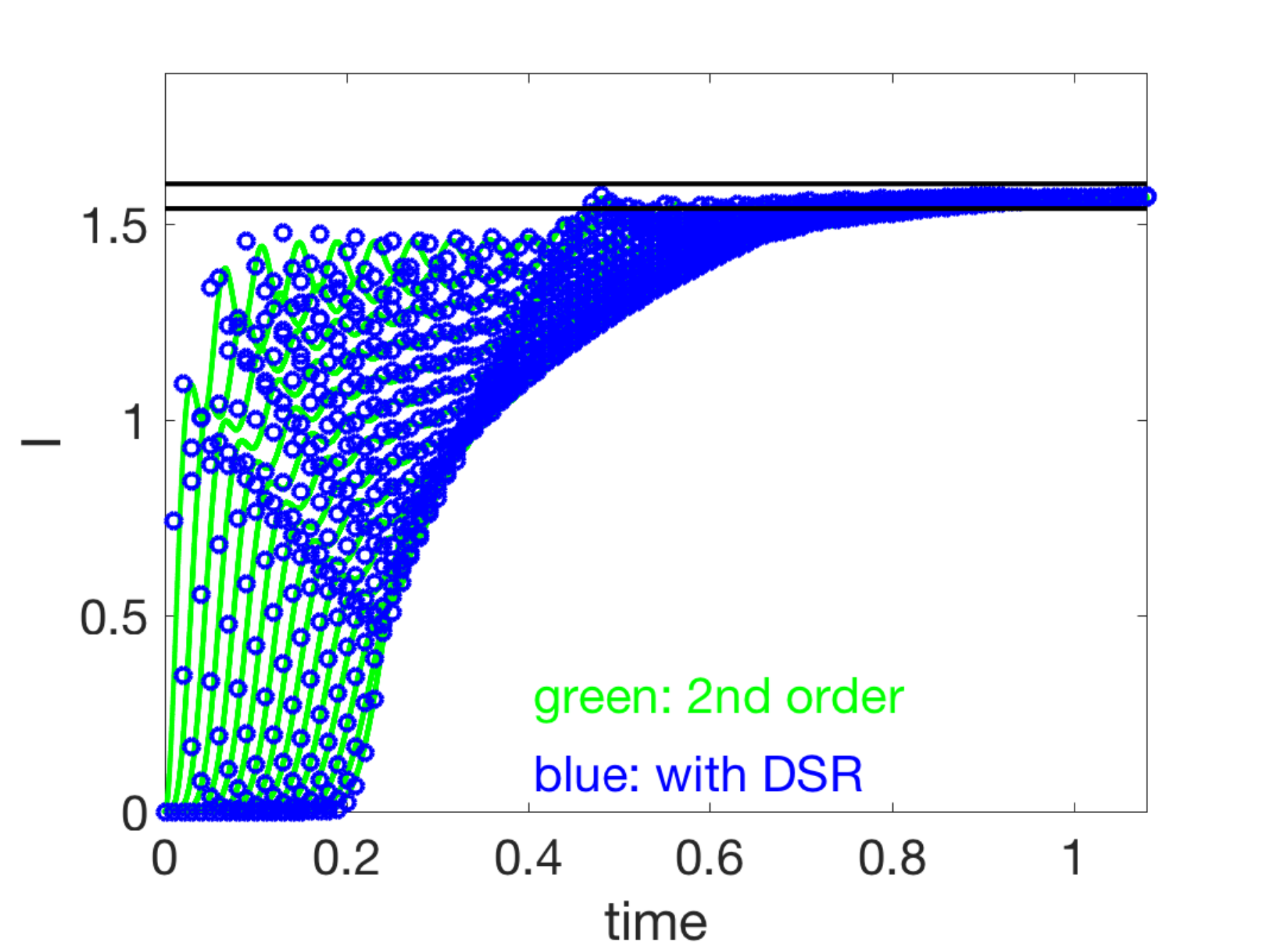} 
  \end{tabular}
   \end{center}
   \vspace{-0.01in}
\caption{
Comparison of step response for (i) the system with DSR and update time $\delta_t= 1 \times 10^{-2}$~s   
and (ii) the second-order system in Eq.~\eqref{system_with_source_2D_tilde} with a 
smaller update time $\tilde{\delta}_t=  8.8759 \times 10^{-5}$~s. 
}
\label{fig_8_step_responses_2ndorder}
\end{figure}

With the same update time $\tilde{\delta}_t = \delta_t = 1 \times 10^{-2}$~s, this system is unstable with a maximum magnitude of eigenvalues $\max_i{\lambda_{\tilde{P},i}} =  1.7667$.  
Reducing the update time by  the scaling factor $ \delta_t \beta$ 
 in Eq.~\eqref{Eq_approximation_DSR_continuous}, i.e.,  $\tilde{\delta}_t = \delta_t ( \delta_t \beta) =  8.8759 \times 10^{-5}$~s 
 results in a stable second-order system dynamics with  a maximum magnitude of eigenvalues $\max_i{\lambda_{\tilde{P},i}} =  0.9995$ and a settling time of $0.9399$~s 
for a  similar  response as that of the system with DSR   as seen in Fig.~\ref{fig_8_step_responses_2ndorder}.  
 Note that a larger update time of   $\tilde{\delta}_t = \delta_t (10 \delta_t \beta) =  8.8759 \times 10^{-4}$~s leads 
 to instability, with  a maximum magnitude of eigenvalues $\max_i{\lambda_{\tilde{P},i}} =  1.0032$.
 Thus, the order-of-magnitude   settling-time reduction with the second-order system also requires an at-least an order-of-magnitude smaller update time. 
In contrast,  the proposed DSR approach enables the faster overall response without the need to use a smaller update time.

\subsection{Impact on ability to maintain  a formation}
\begin{figure}[!h]
\begin{center}
    \begin{tabular}{@{}c@{}}
 \includegraphics[width=0.995\columnwidth]{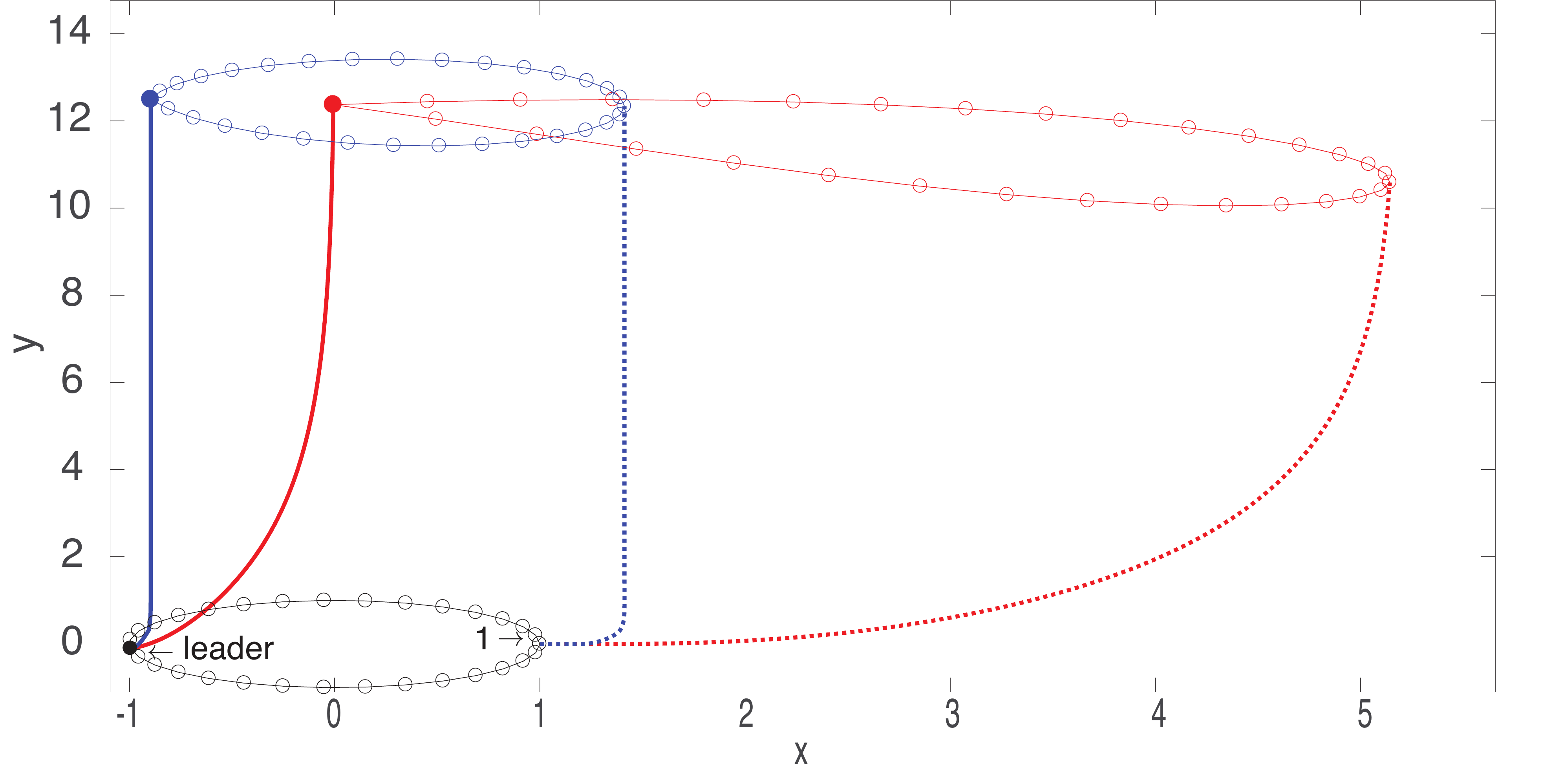} 
  \end{tabular}
   \end{center}
   \vspace{-0.01in}
\caption{
Comparison of initial  (black) and final $x-y$ formation from Eq.~\eqref{eq_formation} with DSR (blue) and without DSR (red). 
Initially, the agents are uniformly arranged on a unit circle centered at the origin shown in black.
The final formation at time $t=12.04$~s 
is shown in red without DSR and blue with DSR. 
The spatial movement of the leader $i^*$ (solid line) and agent $i=n$ (dotted line) are compared with DSR (blue) and without DSR (red).
}
\label{fig_7_compare_xy}
\end{figure}

\vspace{-0.01in}
The impact of the increased response speed with DSR, on the ability to maintain relative positions in a formation (without additional control actions) is also illustrated in the following. 
If the  state $I$ is considered to be the orientation of agents on a plane, then a faster response will allow the agents to 
rapidly coordinating turns and better maintain formations. To illustrate, the $x-y$ position of each agent $i$  was 
computed with the state $I_i$ considered to be the 
orientation of the agent moving with unit speed, as 
\begin{equation}
\begin{array}{rl}
x_i(k+1) = x_i(k) + \delta_t \cos[I_i(k)] \\
y_i(k+1) = y_i(k) + \delta_t \sin[I_i(k)].
\end{array}
\label{eq_formation}
\end{equation}
Initially, the agents are uniformly arranged in a unit circle  in the $x-y$ plane centered at the origin.
Deviations in the orientations $I$ lead to distortions of this initial formation. 
The resulting final formations are compared with and without DSR in Fig.~\ref{fig_7_compare_xy}.

\noindent
Note that, without DSR, the movement of the leader $i^*$  (closest to the information source) is substantially 
different from that of agent $i=n$, which is further away from the information source, which leads to 
substantial distortion of the formation over time. In contrast, the use of DSR reduces the deviations between the agents as seen in Fig.~\ref{fig_6_step_responses},  which 
results in less distortion of the formation, without additional control effort to maintain the formation. Moreover, the smaller settling time implies that the maneuver from initial orientation of $I_i=0$ to 
the final orientation of $I_i = \pi/2$ requires less turn-space to accomplish. In this sense, the proposed DSR can lead to improved 
spatial and temporal response of networked agents, without having to decrease the update time. 

\section{Conclusions}
\vspace{-0.01in}
The delayed self-reinforcement (DSR) method was developed in this article to enable faster network response without the need to increase the individual agent's update rate. 
Stability conditions were developed to facilitate the selection of the DSR-gain parameter. The approach was shown to approximate  a higher second-order 
dynamics for modes corresponding to the smaller eigenvalues of the dynamics. However, the DSR approach does not require the increased update rate 
that would be required for stable operation of such higher second-order dynamics. Simulation results showed that the optimally selected DSR parameters 
were close to the approximation-based estimates of the DSR  parameters. Simulation results also showed that the proposed DSR method  can lead to 
more than an order of magnitude improvement in the settling time.


\end{document}